%% file: elsarticle-template-harv.tex
\documentclass[authoryear,preprint,12pt]{elsarticle}

\usepackage{pdfpages} 
\usepackage{pgffor} 

\makeatletter
\AtBeginDocument{\let\LS@rot\@undefined}
\makeatother

\def\supplementfilename{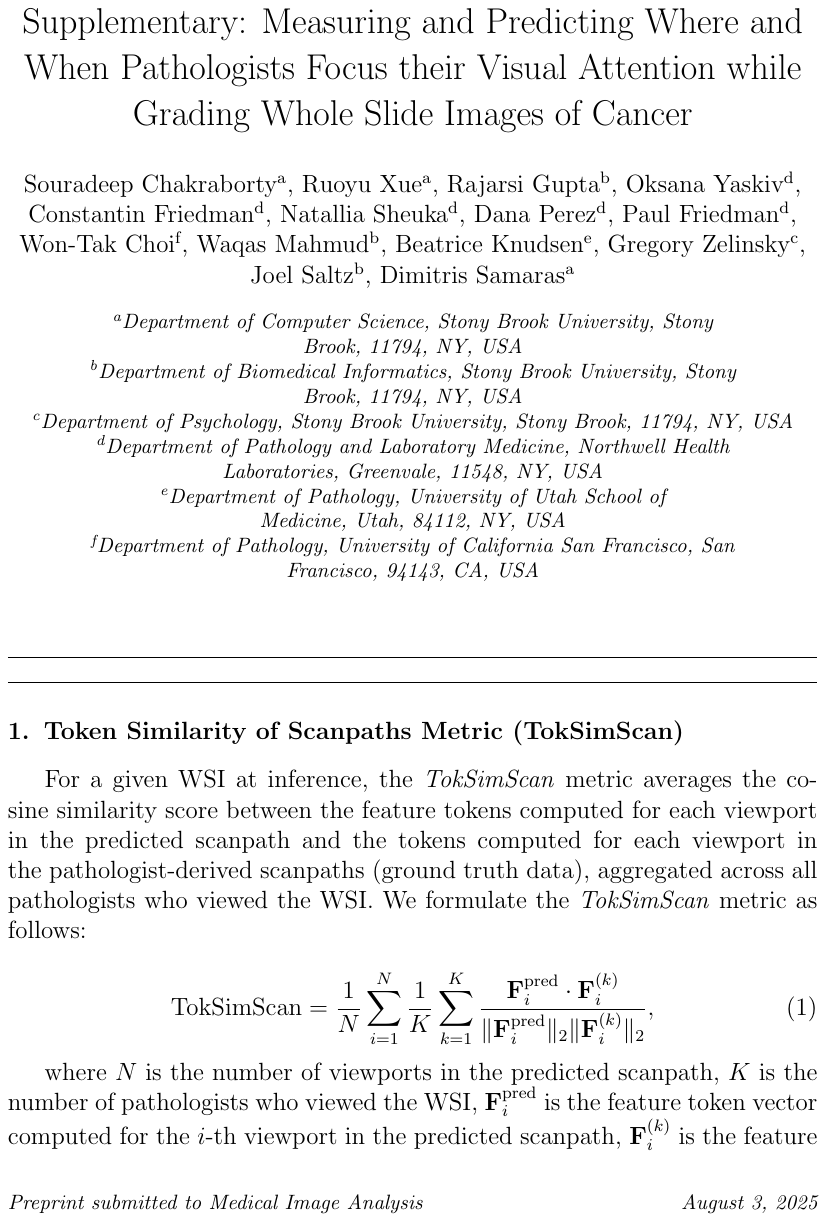}

\pdfximage{\supplementfilename}
\def\numbersupplementpages{\the\pdflastximagepages}

\newif\ifarXiv
\arXivtrue 

\usepackage{amssymb}
\usepackage{amsmath}
\usepackage{xcolor}

\usepackage{lineno}
\usepackage{algorithm}
\usepackage{algorithmic}
\usepackage{booktabs}
\usepackage{subcaption}
\journal{Medical Image Analysis}

\begin{document}

\begin{frontmatter}



\title{Measuring and Predicting Where and When Pathologists Focus their Visual Attention while Grading Whole Slide Images of Cancer} 


\author[lab1]{Souradeep Chakraborty}
\author[lab1]{Ruoyu Xue}
\author[lab2]{Rajarsi Gupta}
\author[lab4]{Oksana Yaskiv}
\author[lab4]{Constantin Friedman}
\author[lab4]{Natallia Sheuka}
\author[lab4]{Dana Perez}
\author[lab4]{Paul Friedman}
\author[lab6]{Won-Tak Choi}
\author[lab2]{Waqas Mahmud}
\author[lab5]{Beatrice Knudsen}
\author[lab3]{Gregory Zelinsky}
\author[lab2]{Joel Saltz}
\author[lab1]{Dimitris Samaras}

\affiliation[lab1]{organization={Department of Computer Science},
            addressline={Stony Brook University}, 
            city={Stony Brook},
            postcode={11794}, 
            state={NY},
            country={USA}}
\affiliation[lab2]{organization={Department of Biomedical Informatics},
            addressline={Stony Brook University}, 
            city={Stony Brook},
            postcode={11794}, 
            state={NY},
            country={USA}}
\affiliation[lab3]{organization={Department of Psychology},
            addressline={Stony Brook University}, 
            city={Stony Brook},
            postcode={11794}, 
            state={NY},
            country={USA}}
\affiliation[lab4]{organization={Department of Pathology and Laboratory Medicine},
            addressline={Northwell Health Laboratories}, 
            city={Greenvale},
            postcode={11548}, 
            state={NY},
            country={USA}}
\affiliation[lab5]{organization={Department of Pathology},
            addressline={University of Utah School of Medicine}, 
            city={Utah},
            postcode={84112}, 
            state={NY},
            country={USA}}
\affiliation[lab6]{organization={Department of Pathology},
            addressline={University of California San Francisco}, 
            city={San Francisco},
            postcode={94143}, 
            state={CA},
            country={USA}}

\begin{abstract}
The ability to predict the attention of expert pathologists could lead to decision support systems for better pathology training.  We developed methods to predict the spatio-temporal (``where" and ``when") movements of pathologists' attention as they grade whole slide images (WSIs) of prostate cancer. We characterize a pathologist's attention trajectory by their x, y, and m (magnification) movements of a viewport as they navigate WSIs using a digital microscope. This information was obtained from 43 pathologists across 123  WSIs, and we consider the task of predicting the pathologist attention scanpaths constructed from the viewport centers. We introduce a fixation extraction algorithm that simplifies an attention trajectory by extracting ``fixations'' in the pathologist’s viewing  while preserving semantic information, and we use these pre-processed data to train and test a two-stage model to predict the dynamic (scanpath) allocation of attention during WSI reading via  intermediate attention heatmap prediction. In the first stage, a transformer-based sub-network predicts the attention heatmaps (static attention) across different magnifications. In the second stage, we predict the attention scanpath by sequentially modeling the next fixation points in an autoregressive manner using a transformer-based approach, starting at the WSI center and leveraging multi-magnification feature representations from the first stage. Experimental results show that our scanpath prediction model outperforms chance and baseline models. Tools developed from this model could assist pathology trainees in learning to allocate their attention during WSI reading like an expert.
\end{abstract}



\begin{keyword}
{Digital pathology  \sep Visual attention  \sep Prostate cancer grading}



\end{keyword}

\end{frontmatter}




\input{sec/introduction_1.tex}
\input{sec/related_works_2.tex}
\input{sec/dataset_3.tex}
\input{sec/methodology_4.tex}
\input{sec/experiments_5.tex}
\input{sec/conclusions_6.tex}
\input{sec/acknowledgments_7.tex}

\bibliographystyle{elsarticle-harv} 
\bibliography{mybibliography}
\clearpage
\ifarXiv
    \foreach \x in {1,...,\numbersupplementpages}
        {
            \includepdf[pages={\x}]{\supplementfilename}
        }
\fi

\end{document}

%% file: sec/introduction_1.tex
\section{Introduction}

The task of reading whole-slide images (WSIs) for cancer diagnosis requires the active collection by attention of cancer-indicating evidence from a WSI, and this highly specialized allocation of attention requires years of training. In radiology, the role of attention during cancer diagnosis has been well documented \citep{gandomkar2016icap,tourassi2013investigating,venjakob2012radiologists,wang2022follow}, and a similar appreciation is now growing in digital pathology \citep{brunye2020eye,brunye2017accuracy,chakraborty2022predicting,chakraborty2022visual,sudin2021eye,chakraborty2024decoding}. Predicting the visual attention of pathologists has the potential to enable development of decision support systems able to guide pathologists as they view and assess whole slide images. The methods we present can potentially also be used to train pathology residents and general pathologists to carry out expert level  sub-specialty interpretations. 

\begin{figure}
\centering
\includegraphics[width = 12.80cm]{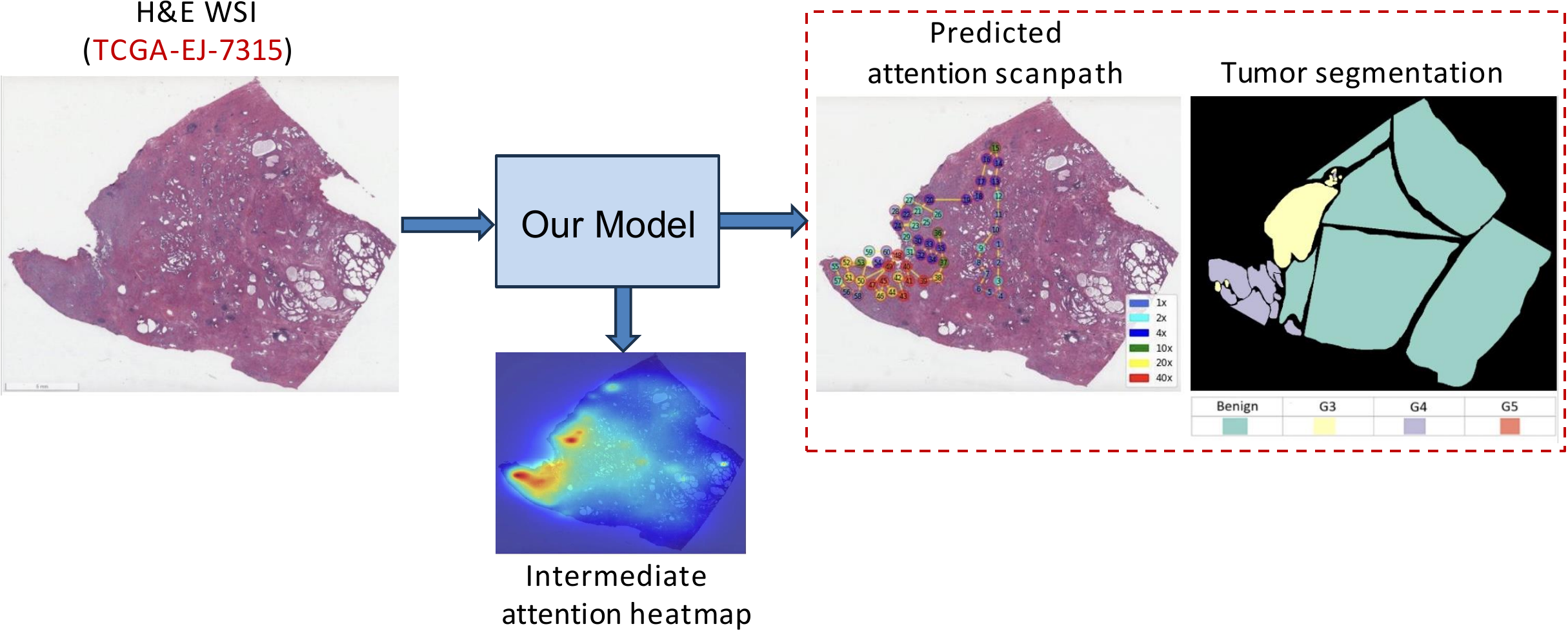}
\caption {Our PAT model predicts attention scanpaths via intermediate attention heatmap prediction for a given WSI (TCGA-EJ-7315) from the TCGA-PRAD dataset.
}
\label{fig:teaser2}
\end{figure}

Previous studies used methods such as eye tracking and mouse movement tracking to investigate pathologists' attention, diagnostic decision-making processes, and expertise-related differences (\cite{bombari2012thinking}, \cite{raghunath2012mouse}, \cite{brunye2017accuracy}, \cite{mercan2018characterizing}, \cite{brunye2020eye}, \cite{sudin2021eye}), and more recently studies have begun to explore predictive models for pathologists' attention during their WSI readings for cancer diagnosis. For instance, in \citep{chakraborty2022visual} we fine-tuned a ResNet34 to predict visual attention heatmaps during prostate cancer grading, and in \citep{chakraborty2022predicting} we employed a Swin Transformer to predict attention patterns during multi-stage gastrointestinal neuroendocrine tumor examinations. Despite these advances, progress has been limited by data scarcity—both in terms of the number of WSIs and participating pathologists. We address this limitation by introducing the largest dataset to date for pathologist attention modeling, comprising 1,016 attention trajectories from 43 pathologists across 11 institutions examining 123 WSIs. This dataset enabled the development of deep learning models that could predict the static visual attention of pathologists (in the form of attention heatmaps) and their expertise levels solely based on how they allocated their attention during their WSI cancer reading. While attention heatmaps provide insights into the spatial distribution of attention, they cannot offer step-by-step guidance to trainees due to their lack of temporal information and this prevents their direct use in computer-assisted pathology training. The ultimate goal is to develop a pathology training tool that guides trainees on where next to attend, and when to make these attention shifts, which would help trainees learn a specialist allocation of attention and potentially reduce inter-observer variability in cancer classifications. Achieving this goal requires predicting the pathologist's attention trajectory.    
Building on our foundational earlier work, here we address the more challenging problem of predicting the spatio-temporal allocation of attention by pathologists performing readings—their ``where" and ``when" allocations of attention. Unlike models that predict the static attention heatmap that focuses on the spatial distribution of attention by collapsing over time, the task of predicting attention scanpaths introduces the additional complexity of modeling temporal dynamics of when attention was allocated to different regions of the WSI. To deal with this added complexity we introduce \textit{Pathologist Attention Transformer (PAT)}, a two-stage model for predicting spatio-temporal attention (scanpaths) of pathologists during their WSI readings. Although this work focuses on prostate cancer grading, the PAT framework is designed to be domain-agnostic and could potentially be extended to other cancer types with sufficient data, as suggested by prior work in gastrointestinal neuroendocrine tumors \citep{chakraborty2022predicting}. As shown for an example WSI in Figure~\ref{fig:teaser2}, our model's prediction of attention scanpath as well as the intermediate attention heatmap align closely with the annotated tumor segmentations (from a Genitourinary specialist), demonstrating its ability to predict the spatio-temporal behavior of pathologists. 

\begin{figure*}[t]
\centering
\includegraphics[width = 13.8cm]{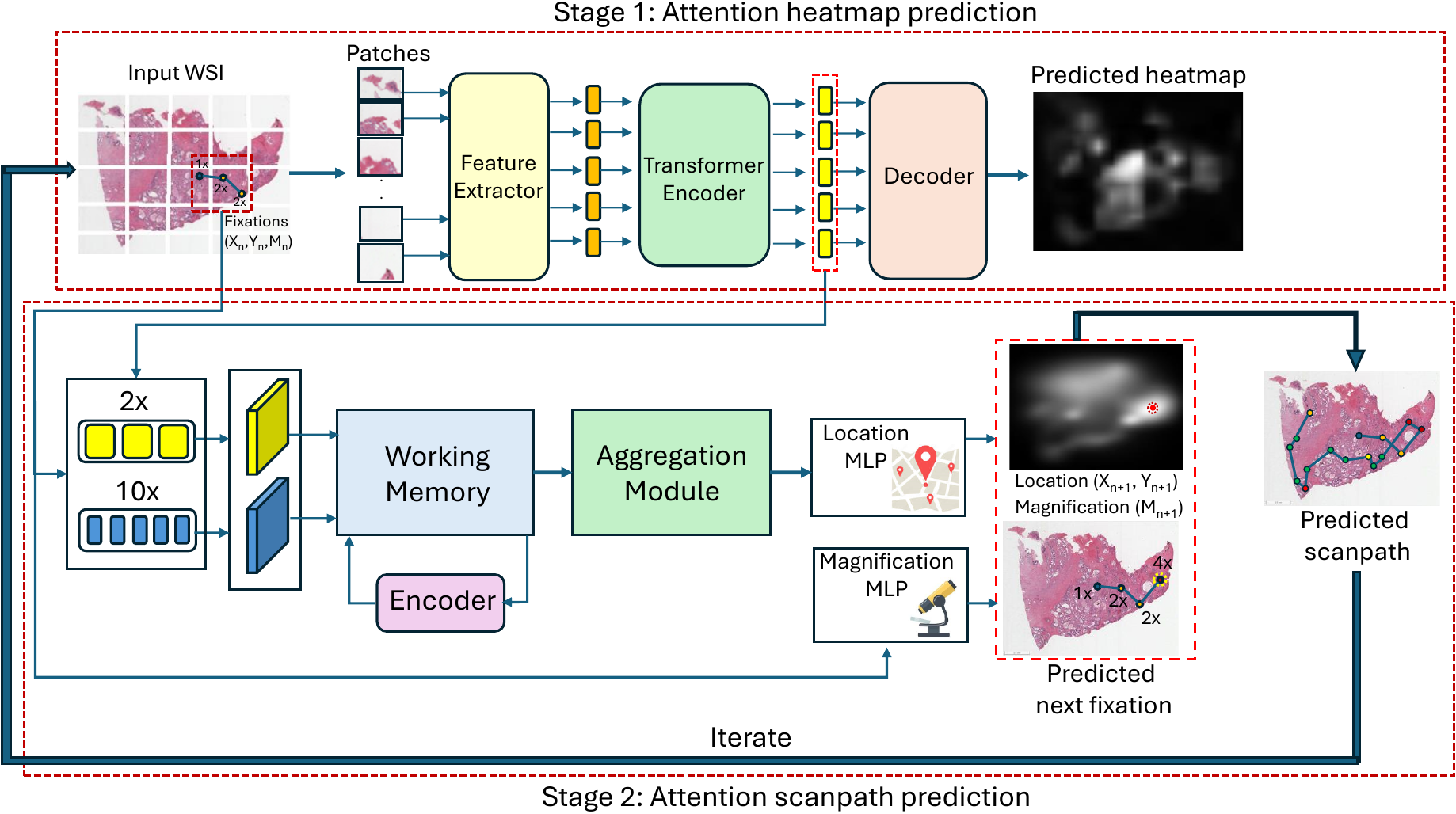}
\caption {Proposed two-stage attention scanpath prediction model, \textit{PAT}. In the first stage, PAT-H predicts pathologists attention heatmap at different magnification levels. In the second stage (PAT-S), we leverage encoded feature representations from this network at low (2X) and high (10X) magnifications to predict the attention scanpath in an autoregressive manner starting from the WSI center. This involves predicting the next fixation $(x,y,m)$ given a sequence of previous fixations, which iterates multiple times to produce the attention scanpath.   
}
\label{fig:proposed_pipeline}
\end{figure*}

Figure \ref{fig:proposed_pipeline} shows the two-stage pipeline of PAT. The first stage, \textit{PAT-Heatmap (PAT-H)}, predicts attention heatmaps for multiple magnification levels using a transformer-based approach. The second stage, \textit{PAT-Scanpath (PAT-S)}, predicts the attention scanpath by combining: 1) a \textit{feature extraction module} that extracts spatial features at low (2X) and high (10X) magnification levels using the encoded feature representations from the first stage, 2) a \textit{foveation module} for dynamically updating working memory with the attended viewport information as well as the information in yet to be attended peripheral regions, 3) an \textit{aggregation module} that serves as a decoder by utilizing a transformer network that selectively aggregates information from the working memory, and 4) \textit{viewport fixation} and \textit{magnification prediction modules} that predict the next viewport fixation location, and the magnification level of this next fixation, respectively. Predicting the next viewport fixation given a pathologist's prior viewing trajectory is a necessary step towards building a training tool capable of giving a trainee pathologist step-by-step guidance about their visual attention at any point in their cancer reading. This process repeats iteratively, enabling scanpath generation in an autoregressive manner starting from the center of the WSI that can be compared to a pathologists' attention during the WSI reading.  While non-autoregressive models are efficient for shorter sequences (e.g., GazeFormer \citep{mondal2023gazeformer}), they lack the ability to capture the sequential dependencies critical for longer scanpaths typical of WSI examinations. In contrast, autoregressive models predict one fixation at a time, dynamically updating their understanding of the visual context. This approach has been shown to improve scanpath prediction performance in the context of undergraduates viewing natural images \citep{yang2024unifying}, and here we extend this iterative autoregressive approach to predict the step-by-step series of decisions made by pathologists conducting WSI readings.

Our study lays the foundation for future AI-assisted pathology training pipelines that can guide trainees on where and how long to focus their attention during WSI readings. By training models on the attention patterns of genitourinary (GU) specialists, we aim to improve the efficiency and diagnostic accuracy of pathology trainees.

In summary, this study makes the following contributions:
\begin{itemize}

    \item We predict the dynamic spatio-temporal attention scanpath of pathologists conducting cancer readings. 

    \item We introduce a two-stage  transformer-based model that predicts the scanpaths of pathologists, and outperforms chance and baseline models. 
    
    \item We propose a novel fixation extraction algorithm that simplifies attention trajectories for model training while preserving semantic context.

    \item We collected the largest known dataset of pathologist attention, comprising 123 WSIs viewed by 43 pathologists from 11 institutions.
\end{itemize}

The paper is structured as follows: Section~\ref{sec:relwork_2} reviews related work, focusing on existing approaches to static and dynamic visual attention modeling, as well as studies on visual attention in digital pathology. Section~\ref{sec:dataset_3} provides a detailed description of our dataset and data processing methods. Section~\ref{sec:methods} outlines the proposed methodology for predicting attention scanpaths using a two-stage transformer-based model. Section~\ref{sec:results} presents the experimental results, analyzing the qualitative and quantitative performance of the proposed methods in attention scanpath prediction. Finally, Section~\ref{sec:conclusion_6} summarizes our findings and proposes potential directions for future research.

%% file: sec/related_works_2.tex
\section{Related Work}
\label{sec:relwork_2}
\subsection{Static Visual Attention (Saliency)}
Traditional approaches to predict visual attention heatmaps or image \textit{saliency} can be categorized as using either a bottom-up \citep{itti1998model,harel2007graph,zhang2008sun,hou2007saliency,zhang2013saliency, chakraborty2016dense} or top-down \citep{yang2016top,kanan2009sun,kocak2014top,ramanishka2017top} modeling approach. Early work by \citep{itti1998model} laid the foundation for bottom-up models by computing contrast between several basic features and using this to predict attention heatmaps of viewers. However, \citep{judd2009learning} highlighted the limitations of a purely bottom-up approach and advocated instead for using higher-level semantic features to improve saliency prediction. Data-driven approaches gained traction with the SALICON dataset by \citep{jiang2015salicon}, which fueled the development of deep learning models that greatly improved the prediction of attention heatmaps. Notable contributions include multi-scale deep features \citep{li2015visual}, multi-contextual features \citep{zhao2015saliency}, and recurrent models \citep{cornia2018predicting}. Recent models have introduced time-specific saliency methods that predict saliency maps in sequential time intervals \citep{aydemir2023tempsal}, and have engaged the problem of inter-observer variability in attention by modeling how individuals shift their focus across diverse visual tasks, thereby paving the way for personalized saliency prediction \citep{chen2024beyond}. See \citep{borji2019saliency,ullah2020brief} for comprehensive reviews of saliency prediction models.

\subsection{Dynamic Visual Attention (Scanpath)}
Attention modeling restricted to saliency map prediction overlooks the temporal dynamics of attentional deployment. Scanpath prediction research can also be characterized as pre- or post-deep learning. For example, \citep{le2015saccadic} showed that adding eye-movement biases (saccade amplitude, orientation, etc.) improved scanpath prediction, and \citep{zanca2019gravitational} modeled the attention scanpath as a movement of a single mass within a gravitational field created by salient visual features. However, most recent work has used deep neural networks incorporating learned semantic features to model attention scanpaths. For example, \citep{kummerer2022deepgaze} proposed DeepGaze III, a framework that integrates a spatial priority network to generate priority maps with a scanpath network conditioned on fixation history. Note that the work reviewed thus far done in the context of a free-viewing task, but there is also an extensive modeling literature predicting attention movements during goal-directed tasks (\cite{zelinsky2020changing, yang2020predicting}). \citep{mondal2023gazeformer} used a natural language model to encode search targets and found that this enabled attention scanpaths to be predicted even in zero-shot contexts where the model was never trained on the target-object category. More recently, \citep{yang2024unifying} introduced a transformer-based architecture HAT that can predict attention scanpaths for both visual search and free-viewing tasks by using a spatio-temporal awareness module akin to the dynamic visual working memory used by humans. While our model shares HAT's autoregressive decoder and working memory design, it introduces key features specific to digital pathology—such as magnification prediction, a two-stage setup with attention heatmap guidance, fixation extraction from raw viewports, and multi-resolution feature integration specific to WSIs. 

These attention models were all built for use with natural images, which have relatively small resolutions (typically a few megapixels or less), and therefore cannot be generalized to giga-pixel WSIs that are much larger in size. WSIs also have a hierarchical structure that requires a multi-resolution analysis to simultaneously capture both global context and fine-grained details. Our study fills this gap by introducing a model of attention prediction that is designed to work with WSIs.

\subsection{Visual Attention in Digital Pathology}
Research into the attention of pathologists has focused on characterizing their eye movements during WSI reading or decoding from eye-movement patterns their level of expertise with a pathology task. Early works analyzed the impact of tumor architecture on prostate cancer grading \citep{bombari2012thinking} and validated the use of mouse cursor movements as a proxy for visual attention during WSI reading \citep{raghunath2012mouse}. Subsequent work studied the relationship between gaze patterns and cancer decision-making, highlighting differences between novices and experts \citep{brunye2017accuracy}. Eye-tracking studies have shown that the gaze behavior of expert pathologists is very efficient and that this results in them having shorter average reading times \citep{warren2018quantifying, sudin2021eye}. For instance, \citep{sudin2021eye} reported fewer fixations and shorter viewing durations among experienced pathologists during breast biopsy interpretation. Much of this work and more can be found in a recent comprehensive review of eye-tracking in digital pathology \citep{lopes2024eye}. Other studies have characterized diagnostic search strategies by viewport tracking, notably scanning (continuous panning at constant zoom) and drilling (zooming through magnifications at different locations) behaviors \citep{mercan2018characterizing}. This study also found that scanning behavior varied with factors such as gender, experience, and institutional setting, though it did not correlate with diagnostic accuracy. 

Despite this excellent start, to date there have been few attempts to model a pathologist's spatio-temporal allocation of attention across changes in magnification during WSI reading. Recent efforts to predict the attention of pathologists trained CNN and transformer-based models on the viewport movements made during WSI readings \citep{chakraborty2022predicting, chakraborty2022visual}, but these works focused on predicting attention heatmaps and did so using limited data available for model training. In very recent work, we used encoders capable of capturing expertise-specific visual patterns in the attention heatmaps of pathologists and used these models to predict their level of expertise \citep{chakraborty2024decoding}. This study leverages our earlier work on attention heatmap modeling and introduces temporal dynamics to predict the attention scanpaths of pathologists. Our work therefore solves the task of dynamic attention prediction of pathologists while also building upon and improving attention heatmap prediction, and by doing so provides a more comprehensive framework for understanding and modeling attention behavior in medical image interpretation.

%% file: sec/dataset_3.tex
\section{Dataset of Pathologist Attention and Cancer Classifications}
\label{sec:dataset_3}
\subsection{Dataset creation}

Similar to ~\citep{chakraborty2022visual}, we used the QuIP caMicroscope, a web-based virtual microscope platform designed for digital pathology \citep{saltz2017containerized}, to collect the attention data and cancer classifications of 43 remotely located pathologists reading WSIs of prostate (TCGA-PRAD dataset) for cancer grading. These pathologists were from 11 different institutions and had expertise levels spanning resident ($n=18$), general ($n=15$), and GU specialist ($n=10$). Upon clicking the link to our research study and reading instruction and consent forms, each pathologist followed the same experimental procedure. A WSI was fit into their viewport (i.e., no magnification) and they were instructed to read the image for Gleason grading while adopting a clinical mindset. To emulate real-world conditions, we did not standardize display specifications such as resolution, monitor size, or color calibration. However, all viewport images were recorded at a fixed resolution of $1050 \times 1680$ pixels, making the data collection agnostic to hardware variability and ensuring consistency for training and evaluation. As they navigated through the WSI in ($x$, $y$, $m$) space, our GUI recorded their $1050\times1680$ viewport image with each mouse-cursor sample (20 Hz). After concluding their reading, the pathologist entered into our interface the primary and secondary Gleason grade and a level of confidence in their decision. This procedure iterated for all the WSI readings in the experiment. 

The 123 WSIs we used for our study were selected by a general pathologist from the 342 prostate WSIs in the TCGA-PRAD dataset \citep{zuley2016radiology}, and the attention data that we collected was processed to obtain heatmaps and scanpaths using methods similar to \citep{chakraborty2022predicting,chakraborty2022visual}. In total, our data collection resulted in 1016 attention scanpaths, 329 from residents, 158 from general pathologists, and 529 scanpaths specialists. On average, each WSI was examined by approximately 8 pathologists and the average reading time per slide per pathologist was 94.68 seconds. Additionally, a GU specialist conducted a grade-level annotation for 22 of the 123 WSI set. 

\begin{algorithm}
\caption{Proposed scanpath simplification algorithm for viewport trajectories}
\label{alg:fix_ext}
\begin{algorithmic}[1]
\renewcommand{\algorithmicrequire}{\textbf{Input: }}
\renewcommand{\algorithmicensure}{\textbf{Output:}}
\REQUIRE Dense scanpath trajectory, $S=\{X_i,Y_i,M_i,T_i\}|_{i=1}^N$  
\ENSURE Simplified scanpath trajectory, $S'=\{X_j,Y_j,M_j\}|_{j=1}^L$
\vspace{0.5em} 

\hrule 
\vspace{0.5em} 

\STATE Split the scanpath $S$ into $R$ scanpath fragments (sub-scanpaths), $\{SF_j\}_{j=1}^R$, each with a constant magnification level $M$.
\STATE Initialize simplified scanpath trajectory, $S' = \{\}$
\FOR {$j = 1$ to $R$}
    \STATE Sub-scanpath, $SS = \{SF_j^p\}_{p=1}^P$   
    \STATE Initialize simplified sub-scanpath trajectory, $SS' = \{SS_1\}$
    \FOR {$p = 2$ to $P-1$}
        \STATE Calculate the angle at point $p$, $A_p = Angle(p)$
        \STATE $T_p^{SS}$ = temporal duration $T$ at index $p$ in $SS$
        \IF {$A_p > Th_A$ and $T_p^{SS} > Th_T$}
            \STATE $SS' = SS' \cup SS_p$
        \ENDIF
    \ENDFOR
    \STATE Sub-scanpath, $SS' = SS' \cup SS_P$
    \STATE Eliminate points from this refined sub-scanpath, $SS'$ based on the dispersion distance between points as:
    \STATE Initialize simplified sub-scanpath trajectory, $SS'' = \{SS'_1\}$
    \STATE Initialize $Temp = T_1$
    \FOR {$q = 2$ to $Q-1$}
        \STATE Calculate the spatial distance with the previous point as:
        \[
        D(q,q-1) = \left \| SS'_q(X,Y) - SS'_{(q-1)}(X,Y) \right \|_2
        \]
        \STATE $T_q^{SS'}$ = temporal duration $T$ at index $q$ in $SS'$
        \IF {$D(q,q-1) \geq Th_D$}
            \STATE $Temp = Temp + T_q^{SS'}$
        \ELSE
            \STATE $SS'' = SS'' \cup SS'_q$
            \STATE $Temp = T_q^{SS'}$
        \ENDIF
    \ENDFOR
    \STATE Sub-scanpath, $SS'' = SS'' \cup SS'_Q$
    \STATE Add the dispersion-distance refined sub-scanpath to the simplified scanpath $S'$ as: $S' = S' \cup SS''$
\ENDFOR
\RETURN Simplified scanpath, $S'$
\end{algorithmic}
\end{algorithm}

\subsection{Scanpath Simplification}
\label{sec:fixext}
Before we can predict a pathologist's scanpath of attention, we first preprocess the viewport movements through the WSI using a method inspired by fixation-extraction algorithms designed to obtain eye fixations from the eye movements while viewing natural images. Our algorithm samples locations along the scanpath trajectory that convey important information about the pathologists' attention, such as where a change in magnification is made, areas viewed for longer durations of time, and points where the scanpath trajectory makes abrupt changes in direction. This scanpath simplification step is important because it decreases data volume, filters out noise, and accelerates analysis, making our simplified attention scanpath more interpretable and noise resistant. We transformed the dense attention scanpath trajectories (sampled every 50 msec) obtained from our caMicroscope interface into simpler attention scanpaths using an in-house scanpath simplification algorithm (Algorithm~\ref{alg:fix_ext}) that constrains the dense trajectory to have at most 150 ``fixation'' points, thus making them more suitable as inputs to predictive models. Another important component of this algorithm is that it retains viewport information wherever and whenever magnification changes, forcing these changes to become fixations in the attention scanpath. During periods of a reading where the magnification remains constant, we use the scanpath simplification algorithm from MultiMatch \citep{dewhurst2012depends} to simplify the pathologist's scanpath. This simplification algorithm involves merging neighboring viewport centers (within a threshold spatial distance), retaining centers with longer viewing durations, and those where sharp turns were made in the scanpath trajectory (where the angle at a point exceeds a threshold). 

 \begin{figure}
\centering
\includegraphics[width = 11.00cm]{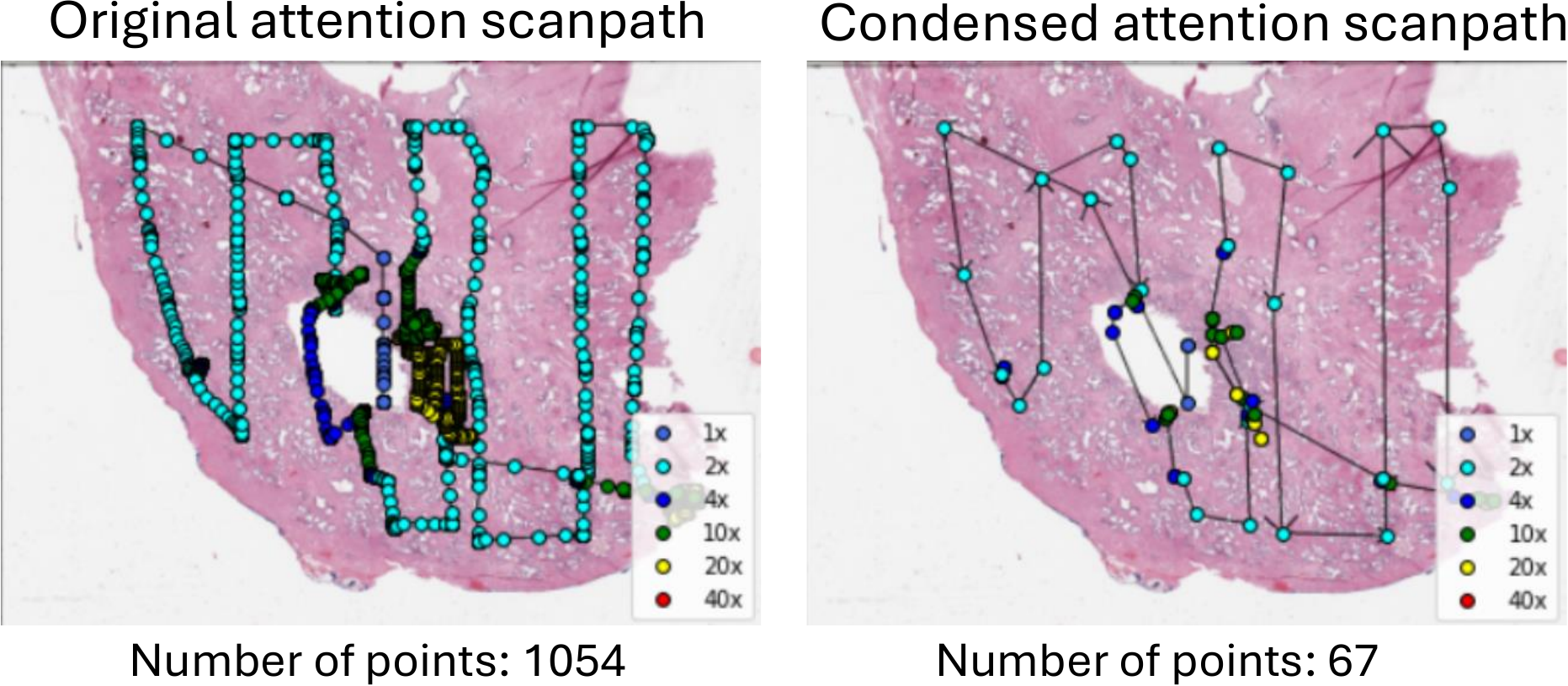}
\caption {Comparison of an original scanpath with a condensed scanpath produced by our scanpath simplification algorithm on case TCGA-2A-A8VL from the TCGA-PRAD dataset.
}
\label{fig:condensed_scanpath}
\end{figure}

Detailed steps for our scanpath simplification algorithm can be found in Algorithm~\ref{alg:fix_ext}, but these steps can be summarized as follows. The algorithm takes as input a dense scanpath trajectory, \( S = \{X_i, Y_i, M_i, T_i\}|_{i=1}^N \), and outputs a simplified trajectory, \( S' = \{X_j, Y_j, M_j\}|_{j=1}^L \) where ($X$,$Y$) is the spatial location of a viewport fixation and $M$ and $T$ denote the magnification and the duration of viewing respectively. The process begins by splitting the scanpath \( S \) into multiple fragments, \( \{SF_j\}_{j=1}^R \), such that each fragment corresponds to a constant magnification level. The simplified trajectory \( S' \) is initialized as an empty set. For each fragment \( SF_j \), a sub-scanpath \( SS \) is processed to generate a simplified sub-scanpath \( SS' \). Simplification involves retaining the first point \( SS_1 \), and iteratively evaluating each intermediate point \( SS_p \) for inclusion based on two conditions: the angle at the point $A_p > Th_A$ and the time spent $T_p^{SS} > Th_T$, where $Th_A$ and $Th_T$ are angular and temporal thresholds. The last point $SS_P$ is always retained. Next, the refined sub-scanpath \( SS' \) undergoes a dispersion-distance refinement. Points are iteratively added to \( SS'' \) based on their spatial distance \( D(q,q-1) \) compared to a threshold \( Th_D \), with temporal information \( Temp \) accumulated for closely spaced points. Finally, \( SS'' \) is updated with the last point \( SS'_Q \) and appended to the global simplified scanpath \( S' \). This iterative process is repeated for all fragments, and the complete simplified scanpath \( S' \) is returned as the final output.

Figure~\ref{fig:condensed_scanpath} illustrates the effectiveness of our scanpath simplification algorithm by visualizing the original and the condensed attention scanpaths obtained using Algorithm~\ref{alg:fix_ext} on a WSI instance from the TCGA-PRAD dataset. Our scanpath simplification algorithm condensed the original scanpath, which was 1054 sampled points, to 67 scanpath fixations. Yet, despite this simplification, the semantic information is largely the same between the two, as seen in the changes in magnification and the similarity in overall global scanpath shape, and these factors make the simplified scanpaths more amenable for training a scanpath prediction model. 

%% file: sec/methodology_4.tex
\section{Methodology}
\label{sec:methods}
\noindent As outlined in Figure~\ref{fig:proposed_pipeline}, we adopt a two-stage method for predicting the dynamic (stage 2) attention of pathologists via intermediate attention heatmap prediction (stage 1). The following subsections describe these two stages in greater detail. 

\subsection{Predicting attention heatmaps}

\begin{figure*}[t]
\centering
\includegraphics[width = 13.8cm]{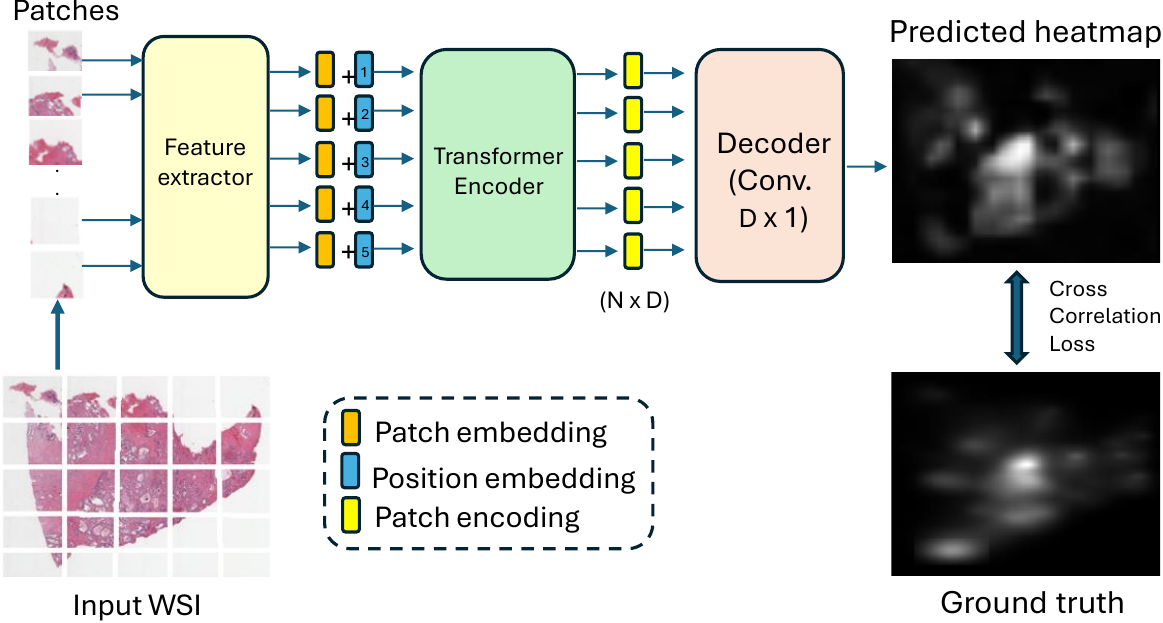}
\caption {Proposed heatmap prediction sub-network of PAT that predicts an attention heatmap for a WSI at different magnification levels. 
}
\label{fig:vis1_heatmap}
\end{figure*}

\noindent Figure ~\ref{fig:vis1_heatmap} shows the pipeline of our heatmap prediction sub-network, built to predict the attention heatmaps obtained from WSI readings.\\
\noindent\textbf{Patch Extraction and Feature Embedding}
Given a WSI \( I \), we split it into a sequence of \( N \) non-overlapping patches, $I = [I_1, I_2, \dots, I_N] \in \mathbb{R}^{N \times P^2 \times C}$, where \( (P, P) \) is the size of each patch, \( N = \frac{H W}{P^2} \) is the number of patches (\( H, W \) are the height and width of the image, respectively), and \( C \) is the number of color channels. Next, we extract patch-wise feature embeddings, $I_0 = [F_{I_1}, F_{I_2}, \dots, F_{I_N}] \in \mathbb{R}^{N \times D}$, where $D$ is the embedding dimension and \( F \in \mathbb{R}^{D \times P^2} \) represents the feature embedding extracted using an off-the-shelf feature extractor, such as ResNet50~\citep{he2016deep},  DINO~\citep{caron2021emerging}, ~\citep{kang2023benchmarking}, etc.\\
\noindent\textbf{Positional Encoding and Transformer Encoder}
To capture positional information, learnable position embeddings, $\text{pos} = [\text{pos}_1, \text{pos}_2, \dots, \text{pos}_N] \in \mathbb{R}^{N \times D}$ are added to the sequence of patch embeddings. This results in the sequence of input tokens $z_0 = I_0 + \text{pos}$. A transformer encoder~\citep{vaswani2017attention} composed of \( L \) layers is applied to \( z_0 \), generating a sequence of contextualized encodings $
z_L \in \mathbb{R}^{N \times D}$.
\\
\noindent\textbf{Decoder and Heatmap Prediction}
The sequence of patch encodings \( z_L \) is decoded into a heatmap \( s \in \mathbb{R}^{H \times W} \) using a convolutional decoder, $
\text{Decoder}: \mathbb{R}^{N \times D} \rightarrow \mathbb{R}^{H \times W}$.
Specifically, a \( D \times 1 \) convolutional layer maps patch-level encodings to patch-level attention scores. The final predicted heatmap \( M_\text{Prd} \) is obtained after normalizing the decoded map.\\
\noindent\textbf{Loss Function}
This network is trained using a loss function based on the cross-correlation (CC) score between the predicted heatmap \( M_\text{Prd} \) and the ground truth heatmap \( M_\text{GT} \). The loss function is defined as:
\begin{equation}
 \mathcal{L} = 1 - \text{CC}(M_\text{Prd}, M_\text{GT}),   
\end{equation}
where the cross-correlation score is computed as:

\begin{equation}
\resizebox{.9\hsize}{!}{
$
\text{CC}(M_\text{Prd}, M_\text{GT}) = \frac{\sum_{i,j} \left(M_\text{Prd}(i,j) - \bar{M}_\text{Prd}\right)\left(M_\text{GT}(i,j) - \bar{M}_\text{GT}\right)}
{\sqrt{\sum_{i,j} \left(M_\text{Prd}(i,j) - \bar{M}_\text{Prd}\right)^2 \sum_{i,j} \left(M_\text{GT}(i,j) - \bar{M}_\text{GT}\right)^2}}
$
}
\end{equation}

Here, \( \bar{M}_\text{Prd} \) and \( \bar{M}_\text{GT} \) denote the mean values of \( M_\text{Prd} \) and \( M_\text{GT} \) respectively, and $i$ and $j$ index the pixels along the width and height of the map. This loss encourages the predicted heatmap to align with the ground truth in terms of both spatial and intensity distributions.

\subsection{Predicting attention scanpaths}

\begin{figure}
\centering
\includegraphics[width = 11.00cm]{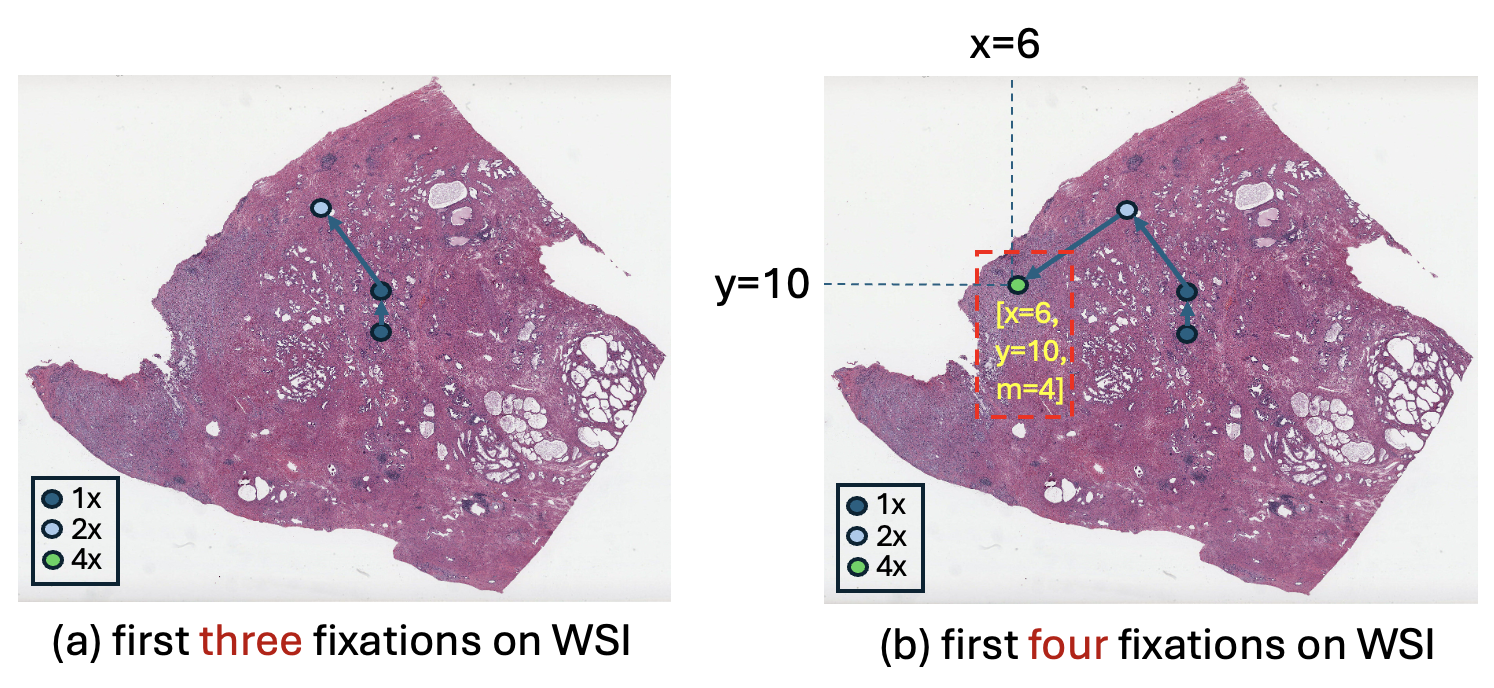}
\caption {Our fixation-by-fixation prediction task for the TCGA-EJ-7315 WSI from the TCGA-PRAD dataset. Our aim is to sequentially predict the next viewport fixation at every step during a pathologist's attention scanpath.
}
\label{fig:vis3}
\end{figure}

\noindent Here we extend our attention heatmap prediction network to the more challenging task of predicting a pathologist's spatio-temporal attention scanpath during a WSI reading. While both non-autoregressive and autoregressive approaches are viable, we opted for an autoregressive model due to its advantages in better handling long fixation sequences and this is likely to be important for capturing the sequential and context-dependent nature of pathologist readings. To capture the iterative decision making that occurs during a pathology reading, we designed our autoregressive model to start at the center of the WSI and to predict each step in the attention scanpath, fixation-by-fixation. In the example illustrated in Figure~\ref{fig:vis3}, given the first three viewport centers (Figure~\ref{fig:vis3}a) the model predicts the fourth viewport center, including its $(x,y)$ location and magnification $m$ (Figure~\ref{fig:vis3}b).  

Inspired by recent work that built a transformer-based model to predict the 
eye fixations in a scanpath \citep{yang2024unifying}, we also use a transformer-based model to predict, in a probabilistic manner, a pathologist's next viewport location and magnification  given a known sequence of prior viewport locations and magnifications. While our model shares the autoregressive decoding framework with HAT \citep{yang2024unifying}, it differs  significantly in scope and design. Unlike HAT, which predicts eye fixations for natural images using (x, y) coordinates, our model addresses clinical scanpath prediction in digital pathology by explicitly modeling viewport-level fixations in (x, y, m) space. In addition, while both models incorporate multi-resolution features, our approach uniquely leverages pathology-specific magnification levels (e.g., low resolution features at 2X and high-resolution features at 10X) and introduces a dedicated magnification prediction module, tailored to model zooming behavior central to WSI reading. Formally, given a WSI $I$ and the prior scanpath trajectory $\mathcal{S}(x_i,y_i,m_i)|_{i=1}^N$ of a pathologist as inputs, the model outputs the next viewport  $\mathcal{S}(x_{(N+1)},y_{(N+1)},m_{(N+1)})$ at every step, where $(x,y)$ denotes the spatial location of the viewport center in the map and $m$ denotes the magnification of the viewport.

\begin{figure*}
\centering
\includegraphics[width = 13.8cm]{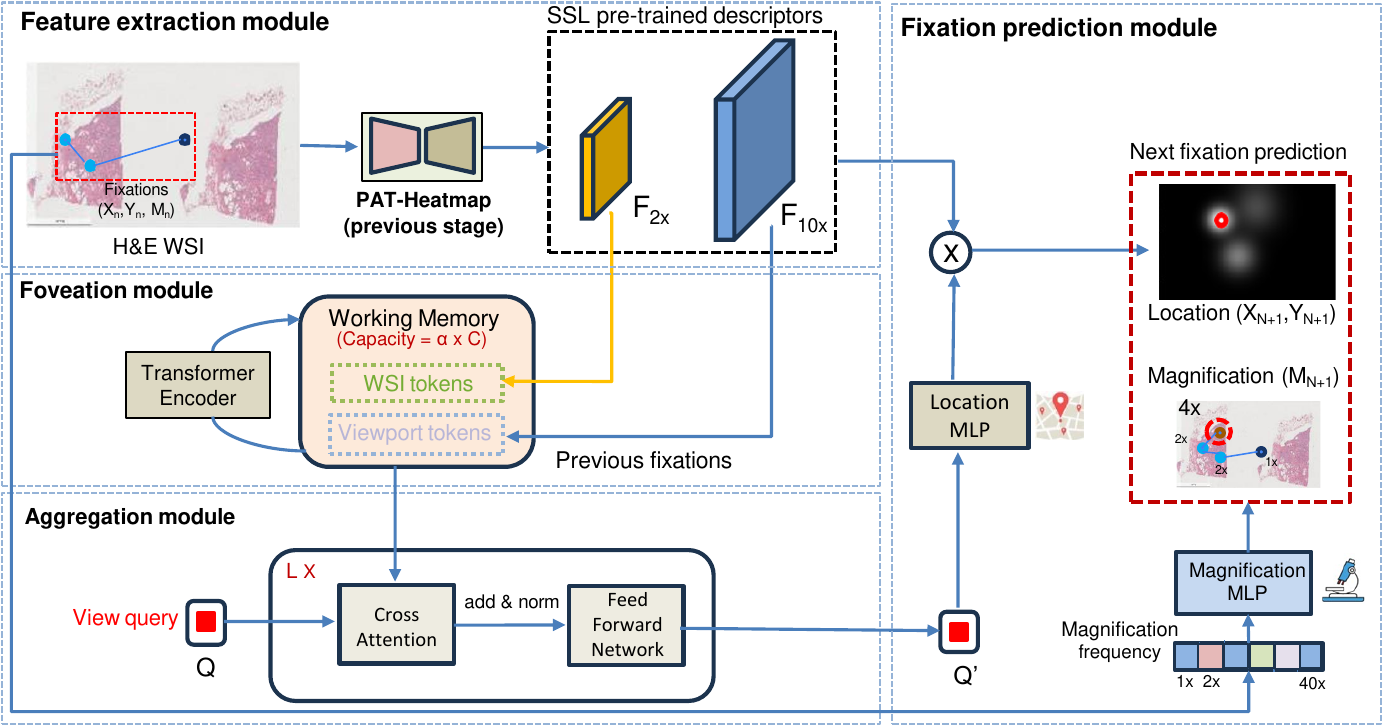}
\caption {The proposed PAT-Scanpath sub-network  predicts the next viewport (location and magnification) of a pathologist on a WSI based on their prior scanpath trajectory and the WSI as inputs. Encoded feature embeddings at low (2X) and high (10X) magnification levels from our PAT-Heatmap (stage 1) are utilized to construct the feature space. A working memory with a capacity of 
$\alpha$ tokens is formed by combining feature vectors from $F_{2X}$ with those of $F_{10X}$ at previously fixated locations, representing both WSI-wide and viewport-specific information. A transformer encoder dynamically updates this working memory at each new fixation. The model then generates a single query vector of dimension $C$, which aggregates information from the shared memory to predict fixations. Finally, the updated query is convolved with $F_{10X}$ through an MLP layer to produce fixation heatmaps, while magnification levels are predicted through a separate MLP layer.
}
\label{fig:PAT-S_pipeline}
\end{figure*}

Figure~~\ref{fig:PAT-S_pipeline} shows the pipeline of the stage 2 sub-network of our PAT model for predicting the next viewport fixation (location and magnification) in an attention scanpath given the prior scanpath trajectory as an input. Following work that predicted scanpaths of eye fixations from people viewing natural images \citep{yang2024unifying}, we designed our PAT model to have four functionally different modules that act in sequence: 1) a feature extraction module that directly leverages multi-resolutional feature encodings at different magnifications from our PAT-Heatmap sub-network (stage 1), 2) a foveation module that maintains a dynamical working memory representing the information acquired through viewport fixations over time, 3) an aggregation module that selectively aggregates the information in the working memory using attention mechanisms, 4) a viewport fixation prediction module that predicts the fixation heatmap $H$, and 5) a magnification prediction module that predicts the magnification level $m$ of the next fixation. The following are more detailed descriptions of each module.\\
\noindent\textbf{The feature extraction module} gathers feature encodings across multiple magnifications from the PAT-Heatmap network (our stage 1 sub-network for heatmap prediction) and assembles these encodings into 3D feature maps, $F_{2X}$ and $F_{10X}$ for each of the low (at 2X) and the high magnification levels (at 10X), respectively. While the multi-scale design of \citep{yang2024unifying} uses a feature pyramid to simulate foveated human vision during natural image viewing—where high-resolution regions represent visual focus and lower-resolution areas mimic peripheral context—our use of multi-resolution features serves a different purpose. In the context of digital pathology, magnification levels (e.g., 2X vs. 10X) are not perceptual approximations but clinically meaningful scales that pathologists explicitly select to examine different structural or morphological features. Thus, our multi-resolution design reflects the clinically relevant diagnostic reasoning process, where different magnifications reveal distinct semantic content critical for grading cancer.\\
\noindent\textbf{The foveation module} constructs a \textit{dynamic} working memory tailored for WSI reading by combining multi-resolution features from both unexplored and previously attended regions. Specifically, we use feature maps from 2X and 10X magnifications: the low-resolution map $F_{2X}$ provides \textit{WSI tokens} representing information from yet-to-be-visited regions, while the high-resolution map $F_{10X}$ provides \textit{viewport tokens} representing information from prior fixation locations.

To form the working memory with $\alpha$ tokens, we flatten $F_{2X}$ spatially to extract WSI embeddings, and select viewport embeddings from previous fixations in $F_{10X}$. A transformer encoder updates this memory with each new fixation. As in HAT \citep{yang2024unifying}, we incorporate spatial position encodings, scale embeddings, and temporal embeddings into the token representations. However, in contrast to HAT’s foveated vision simulation for natural image viewing, our tokens reflect clinically meaningful magnifications—where 2X and 10X correspond to diagnostically distinct perspectives (e.g., tissue architecture vs. cellular morphology).

Importantly, we extend the token representations with a learnable \textit{magnification embedding} to encode the magnification level at which each viewport was observed. This addition reflects the explicit and interpretable role of magnification in pathology decision-making, which is not present in attention modeling in natural images.

\noindent\textbf{The aggregation module}, adapted from the autoregressive decoder in HAT \citep{yang2024unifying}, is a transformer-based decoder that aggregates contextual information from the working memory using a learnable query vector $Q \in \mathbb{R}^{1 \times C}$. At each decoding step, $Q$ attends to the memory via cross-attention, followed by a feed-forward transformation to produce the updated query representation $Q'$. This process is repeated over $L$ decoder layers. Unlike HAT, which includes both cross-attention and self-attention to model interactions across multiple task-specific queries, our model is designed for a single-task setting (prostate cancer grading) with a shared query across all decoding steps. Therefore, self-attention across queries is not necessary. Temporal context is fully captured via the evolving working memory, enabling a more efficient decoder while retaining the ability to model sequential viewing behavior.

\noindent\textbf{The fixation prediction module} is conceptually  similar to the fixation prediction module in HAT \citep{yang2024unifying} and predicts the attention heatmap $\hat{H}$ using a Multi-Layer Perceptron $MLP_H$ having two hidden layers.  $MLP_H$ first transforms the query $Q'$ into an embedding, and then convolves this embedding with the high-resolution feature map $\mathcal{F}_{10X}$ to get the fixation heatmap $\hat{H}$ after a sigmoid layer:
\begin{equation}
\label{eq:heatmap}
    \hat{H}=\text{sigmoid}(\mathcal{F}_{10X} \odot {MLP_H}(Q'))
\end{equation}
where $\odot$ denotes the pixel-wise dot product operation. Finally, we upsample $\hat{H}$ to the image resolution.\\
\noindent\textbf{The magnification prediction module} is a novel component of our model, designed to capture the zooming behavior of pathologists—an aspect not modeled in prior scanpath prediction methods such as HAT \citep{yang2024unifying}. Unlike natural image viewing, pathologists explicitly adjust magnification to examine tissue at different scales, making magnification prediction crucial for realistic WSI scanpath modeling. We first compute the cumulative magnification count \( CM \in \mathbb{R}^{1 \times M} \) for the input scanpath \( S \) as:
\begin{equation}
CM = \{ CM_r \}_{r=1}^M, \quad \text{where} \quad CM_r = \sum_{v=1}^N \mathbb{I}[m_v = r]
\end{equation}
Here, $\mathbb{I}[m_v = r]$ is an indicator function that equals 1 if \( m_v = r \), and 0 otherwise. \( r \) represents the magnification levels indexed in the list of magnifications \([1X, 2X, 4X, 10X, 20X,40X]\). \( CM \) is a vector of length \( M \) representing the frequency of each magnification level. For example, if the sequence of viewport magnifications is \([1X, 1X, 2X, 2X, 2X, 4X, 10X, 10X]\), the output vector corresponding to the magnification levels \([1X, 2X, 4X, 10X, 20X, 40X]\) will be \([2, 3, 1, 2, 0, 0]\). Instead of directly passing the actual sequence of magnifications of previous viewport fixations that contains noisy frequency magnification transitions, we rather count the frequency of the magnifications until the fixation index $N$, and consider this 6-dimensional vector (for each of the 6 magnifications) as our magnification feature descriptor. Next, we pass this descriptor $CM$ through an MLP layer to predict the magnification level, which is a 6-way classification task. For the MLP, a linear layer followed by a sigmoid activation is applied on top of the vector $CM$ to predict magnification level $\hat{m}$:
\begin{equation}
    \hat{m} = \text{sigmoid}(W \cdot {CM}^T+b),
\end{equation}
where $W$ and $b$ are the parameters of the linear layer.\\

\noindent \textbf{Training} We use behavior cloning to train our model following \citep{zelinsky2019benchmarking,yang2024unifying}. We decompose the problem of scanpath prediction into learning a mapping from the input pair of an image and a sequence of previous fixations to the output pair of a fixation heatmap and magnification level. Given the predicted fixation heatmap $\hat{Y}\in\mathbb{R}^{H{\times} W}$ and magnification level  $\hat{m}\in\mathbb{R}^{1{\times}1}$, the training loss is calculated as:
\begin{equation}
    \mathcal{L}=\mathcal{L}_\text{fix}(\hat{Y}, Y)+ \lambda_{mag}\mathcal{L}_\text{mag}(\hat{m}, m),
\end{equation} 
where $Y\in[0,1]^{H{\times} W}$ and $m\in\{1,6\}$ are the ground-truth fixation heatmap and magnification level, respectively and  $\lambda_{Mag}$ is the parameter for the magnification classification loss. We compute $Y$ by smoothing the ground-truth fixation map with a Gaussian kernel having a kernel size inversely proportional to the magnification level. Thus, the lower the magnification level, the higher the kernel size. $\mathcal{L}_\text{fix}$ denotes the fixation loss and is computed using pixel-wise focal loss \citep{lin2017focal,law2018cornernet,yang2024unifying}:
\begin{equation}
\begin{aligned}
\mathcal{L}_\text{fix} = \frac{-1}{H'W'}\sum_{i,j}
\begin{cases} 
    (1-\hat{H}_{ij})^\gamma\log(\hat{H}_{ij}) & \textrm{if } H_{ij}=1,\\[5pt]
    \begin{gathered}
    (1-{H}_{ij})^\beta(\hat{H}_{ij})^\gamma\\
    \log(1-\hat{H}_{ij}) 
    \end{gathered}& \text{otherwise},
\end{cases}
\end{aligned}
\label{eq:loss_fn}
\end{equation}
where $H_{ij}$ represents the value of map $H$ at location $(i, j)$, H' and W' are the height and width of the output high-resolution density map, and we set $\gamma=2$ and $\beta=4$ following \citep{yang2022target,law2018cornernet}.

$\mathcal{L}_\text{mag}$ is the magnification loss (6-way classification for the 1X, 2X, 4X, 10X, 20X, 40X magnification levels) and is computed by applying a weighted cross entropy (negative log-likelihood) loss, i.e., 
\begin{equation}
    \mathcal{L}_\text{mag} = - \sum_{c=1}^{C} w_c \, m_c \log(\hat{m}_c),
\end{equation}
where $m_c$ and $\hat{m}_c$ are the ground truth and predicted magnification levels, and $w_c$ is the weight corresponding to the magnification level $c$. 

\begin{equation}
w_c = \frac{N}{C \cdot N_c}
\end{equation}
where, $N$ is the total number of samples in the dataset, $C$ is the total number of classes, and $N_c$ is the number of samples in class c. This weighting ensures that the model assigns appropriate importance to different classes during training, addressing issues like class imbalance.\\

\noindent\textbf{Inference} During inference, the next fixation location is deterministically selected from the predicted attention heatmap using the \textit{argmax} rule, as this method demonstrated superior performance compared to probabilistic sampling. In contrast, we employ a probabilistic sampling strategy for magnification prediction rather than a deterministic approach. This decision stems from observed high inter-observer variability in magnification transitions. Empirically, probabilistic sampling of the magnification level from the predicted class logits proved more effective than deterministic methods, as it better captured the inherent variability (see Section~\ref{sec:ab_study} for detailed ablation studies).

To simplify magnification transitions, we assume that the magnification level can only increase, decrease, or remain unchanged relative to the magnification of the last fixation. Formally, let \( m_t \) denote the magnification of the last viewport at time step \( t \), and let \( \mathbf{p} = [p_{1X}, p_{2X}, \ldots, p_{40X}] \) represent the probabilities (logits) corresponding to each magnification class. The predicted magnification \( \hat{m}_{t+1} \) for the next fixation is computed as:
\begin{equation}
\hat{m}_{t+1} = m_t + \operatorname{argmax}_{\Delta \in \{-1, 0, 1\}} p_{m_t + \Delta},
\end{equation}
where \( p_{m_t + \Delta} \) is the probability of transitioning to the magnification \( m_t + \Delta \), and \(\Delta \in \{-1, 0, 1\}\) represents the possible transitions: decrease, no change, or increase. For example, if the magnification of the last fixation \( m_t \) is \( 2X \), and the predicted logits are 
\([p_{1X} = 0.05, p_{2X} = 0.10, p_{4X} = 0.30, p_{10X} = 0.20, p_{20X} = 0.30, p_{40X} = 0.05]\), 
the next magnification is calculated as \( \hat{m}_{t+1} = 4X \), since \( p_{4X} = 0.30 \) is the highest probability among the feasible transitions. While one could model magnification transitions using a 3-way classification over $\Delta \in \{-1, 0, 1\}$ adding the predicted shift to the current magnification level, we instead predict full 6-way logits over the magnification levels 1X--40X and constrain the magnification transition during inference. This design captures clinically meaningful differences between magnification levels and provides richer supervision (more fine-grained feedback and stronger gradients) during training, allowing the model to learn finer-grained patterns in pathologist behavior. As shown in our ablation (see supplementary), this leads to improved scanpath prediction performance over the simpler direction-only approach. 

We iteratively predict all viewport fixations to generate the entire scanpath in an auto-regressive manner, taking the WSI \( I \) and the first fixation at the center of the WSI as inputs. At each step \( i \), the proposed PAT-Scanpath predicts the center coordinates \((x_i, y_i)\) of the next viewport and magnification \(m_i\), producing the full scanpath 
\(\mathcal{S} = \{(x_i, y_i, m_i)\}_{i=1}^N\) by iterating until \(N\) fixations are generated. The inference scanpath length, $N$ is determined by the average sequence length in the training set. Please refer to the supplementary material for detailed ablations on the choice of $N$. This process is described as:
\begin{equation}
\mathcal{S} = \operatorname{PAT-S}(I; (x_1, y_1, m_1)),
\end{equation}
where \((x_1, y_1, m_1)\) represents the initial viewport parameters ($x_1,y_1,m_1 = \frac{H'}{2},\frac{W'}{2},1X$) at the WSI center at ($\frac{H'}{2}$,$\frac{W'}{2}$). This sequential approach ensures that each predicted viewport dynamically depends on its previous predictions, aligning with the context-aware nature of human visual attention during WSI reading. At each step, we apply Inhibition-of-Return (IOR) on the predicted attention heatmap, following existing models \citep{navalpakkam2005modeling,tatler2005visual}. This step suppresses revisits to recently attended WSI regions (i.e. locations already in the prior scanpath), thereby enhancing exploratory visual behavior.  

%% file: sec/experiments_5.tex
\section{Experiments}
 \label{sec:results}
 In this section, we present qualitative and quantitative evaluations of the predictive success of our PAT model, and compare these predictions to those from baseline models for attention scanpath prediction.
 
We adopted 5-fold cross-validation for evaluation,  randomly partitioning our dataset of 123 WSIs into five folds (four folds with 25 WSIs each and one with 23 WSIs). All models were trained on four folds and evaluated on the remaining fold, using identical train/test splits for both the heatmap prediction (stage 1) and scanpath prediction (stage 2) steps. Splitting was performed strictly at the WSI level, such that all attention trajectories and annotations associated with a given WSI appeared only in a single fold. Although pathologists contributed readings across multiple WSIs, no explicit separation by pathologist or institution was enforced, as attention behavior in pathology is primarily image-driven. Additionally, the feature encoders (e.g., DINO) used during model training were pretrained externally on natural images and kept frozen, ensuring that training relied solely on downstream task data without any use of test WSI statistics.

\subsection{Evaluation Metrics}
Our evaluation of the scanpath prediction models takes a two-pronged approach, asking: 1) how similar the predicted scanpaths are to the
pathologist-derived scanpaths, and 2) how accurately the model predicts the next viewport given the history of previous viewport fixations in the attention trajectory. 

For scanpath similarity metrics, we use: 1) the averaged NSS score \citep{ohlschlager2017scegram}, 2) the averaged AUC \citep{judd2009learning} score, following existing literature \citep{kummerer2022deepgaze}, 3) Semantic Sequence Score (SSS) \citep{chakraborty2022visual}, which measures the average similarity between the sequences of cancer semantic segmentations underlying the viewport fixations in the predicted scanpath and those in the pathologist-derived scanpaths, and 4) the average token similarity (\textit{TokSimScan}) between the viewports in the predicted scanpath and those in the pathologist-derived scanpaths across different magnification levels.

Existing metrics for evaluating scanpath prediction primarily focus on sequence-based comparisons. For instance, the Sequence Score (SS) metric \citep{borji2013analysis} evaluates the similarity of fixation-based clusters, while the Semantic Sequence Score (SSS) metric \citep{chakraborty2022visual,yang2022target} compares sequences of semantic labels. However, no existing metric measures the similarity of feature tokens corresponding to fixations in predicted and ground truth scanpaths. To address this gap, we introduce  the \textit{TokSimScan} metric, which quantifies the similarity between feature tokens of predicted viewport fixations and those derived from pathologist scanpaths for each magnification level. See the supplementary material for the formal definition of TokSimScan and a detailed discussion of the motivation behind its design. The Semantic Sequence Score (SSS) metric measures inter-observer scanpath similarity between the predicted scanpath and the pathologist-derived scanpaths (ground truth), specifically in terms of the grades of tumor regions traversed during WSI viewing. Following \citep{chakraborty2022visual}, we derived SSS by adapting the Sequence Score (SS) metric \citep{borji2013analysis}, originally designed to compare scanpaths on natural images, by replacing clusters based on eye fixations with Gleason-graded regions (derived from tumor segmentation annotations) at the viewport fixations. While \citep{chakraborty2022visual} used this metric to measure similarity between the scanpaths of two pathologists, we used it to measure the similarity between the predicted scanpath and the pathologist scanpaths. In this approach, each scanpath is converted into a string that represents the sequence of Gleason grades corresponding to the viewport centers (e.g., $B-G_3$-$G_5$-$G_4$-$G_4$, $B-B-G_4$-$G_4$-$G_3$-$G_3$-$G_5$, etc., where $B$ denotes benign regions, and $G_n$ denotes Gleason grade $n$). A string-matching algorithm \citep{needleman1970general} is then applied to quantify the similarity between these grade sequences. 

For our evaluation consisting of predicting the next viewport given a history of previous viewports, we compare model performances using: 1) the normalized Euclidean distance between the predicted and the ground truth next viewport fixation location, 2) average token similarity of the predicted and the ground truth next viewport fixations (we call this \textit{TokSimFix}), 3) magnification prediction accuracy (\%) across the different magnification levels, and 4) accuracy of predicting magnification change (\%) across the different magnification levels.

\subsection{Baselines}
We compare the performance of our model against different baseline models - 

1) \textit{Random1}: a chance baseline that randomly predicts both location $(x,y)$ and magnification $m$ of the next viewport fixation, 

2) \textit{Random2}: another chance baseline that uses the location and magnification of the viewport based on the attention data of a randomly selected pathologist on a different WSI, also selected at random (excluding the test WSI). For evaluating the performance on predicting the next viewport fixation, we select a viewport fixation on an attention scanpath from the same pathologist but viewing a different WSI at the same fixation number. For the scanpath prediction task, we select the scanpath of a pathologist selected at random on a different WSI and assign it as the predicted scanpath, 

3) \textit{VanFormer}: a vanilla Transformer model \citep{vaswani2017attention} trained to predict the location and magnification of the next viewport,  $(\hat{x}_{N+1},\hat{y}_{N+1},\hat{m}_{N+1})$ directly based on the prior sequence of viewport fixations  $(x_i,y_i,m_i)|_{i=1}^N$. Unlike our PAT model, which predicts magnification probabilities and intermediate heatmaps, this model directly outputs the exact location ($x,y$) and magnification ($m$) values for the next viewport fixation. We select this model as a baseline because Transformer models have proven effective in processing sequential data \citep{vaswani2017attention} and predicting subsequent values due to their self-attention mechanism that captures complex dependencies within the input sequence,

4) \textit{VanSemFormer}: an extension of the vanilla Transformer model that additionally takes as input the feature token information $t$ thus forming the input $(x_i,y_i,m_i,t_i)|_{i=1}^N$ for all viewport fixations in the scanpath sequence (of length $N$), 

5) \textit{GazeFormer} \citep{mondal2023gazeformer}, that predicts the entire scanpath in a non-autoregressive manner, i.e., generating all viewports in a single step rather than sequentially. For training this model, we utilized the 10X feature maps derived from our PAT-H sub-network, as 10X magnification is the most commonly employed level for prostate cancer grading. The baseline models VanFormer, VanSemFormer, and GazeFormer were trained using the Mean Absolute Error loss (or L1-loss) between the predicted 3-tuple $(x_{N+1},y_{N+1},m_{N+1})$ vector (predicting location and magnification) with the corresponding ground truth location and magnification $(\hat{x}_{N+1},\hat{y}_{N+1},\hat{m}_{N+1})$. See the supplementary for more implementation details.

\noindent We compare two different versions of our PAT  model --
\begin{enumerate}
\item The ``PAT-PriorMag" model takes a Bayesian approach to predict the magnification level, where the magnification level is randomly selected based on the prior probability of magnification transitions in the training data.
\item The ``PAT-ProbMag" model implements the  inference approach discussed in Section~\ref{sec:methods} by probabilistically determining the direction of magnification change (increase or decrease), which is then added to the current magnification level to predict the magnification level of the next fixation.
\end{enumerate}

\textbf{Feature Encodings.} To evaluate the impact of different visual representations, we experimented with four types of feature encodings for our PAT model: (1) \textit{DINO-Vanilla}, i.e., off-the-shelf DINO features pretrained on ImageNet; (2) \textit{Kang-Vanilla}, i.e., features from the histopathology-pretrained model in \citep{kang2023benchmarking}; (3) \textit{DINO-PAT-H}, i.e., features extracted from our PAT-H sub-network trained using DINO-Vanilla features; and (4) \textit{Kang-PAT-H}, i.e., features from the PAT-H sub-network trained using Kang-Vanilla features. Our PAT-H sub-network (Stage 1) can be trained on either DINO or \textit{Kang} features, and its output feature encodings are then reused for our  downstream scanpath prediction task. We evaluated the \textit{Kang} features for this task because their pretraining on large-scale digital pathology datasets makes them domain-specific, and thus better suited for capturing histopathological patterns compared to generic self-supervised features.

\subsection{Results}
\begin{figure*}
\centering
\includegraphics[width = 13.8cm]{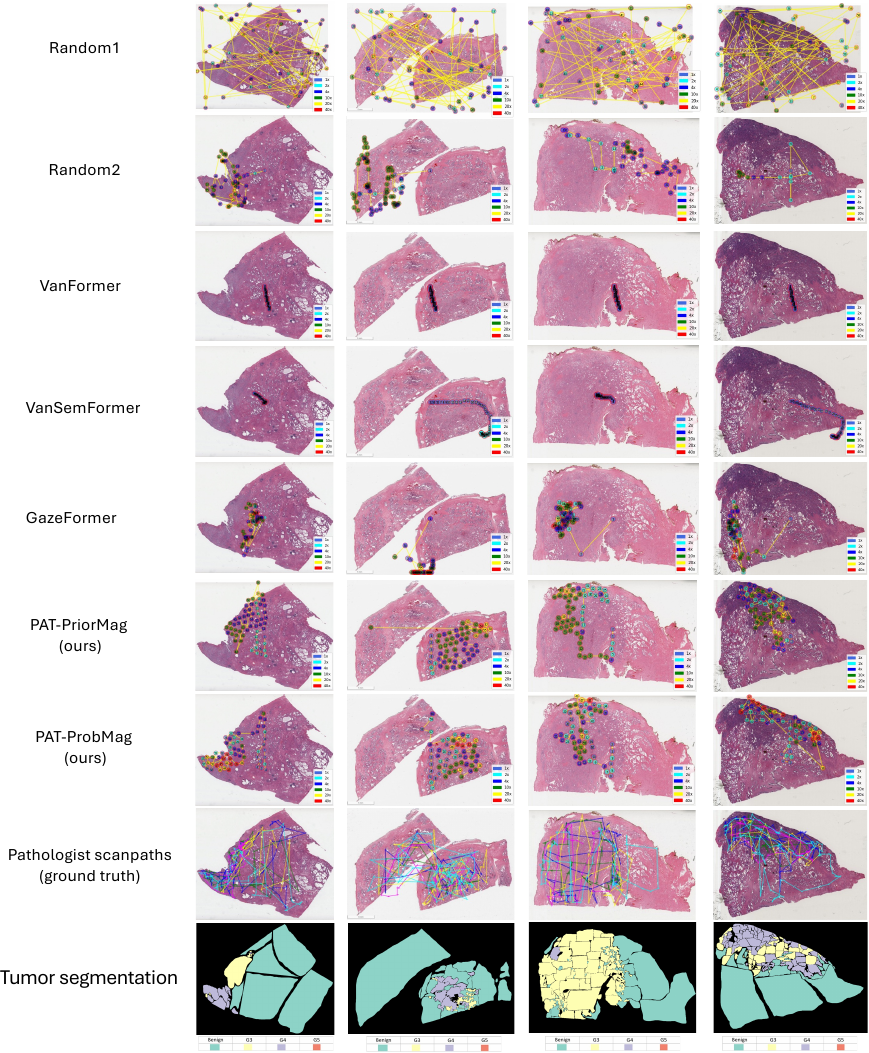}
\caption {Qualitative comparison of attention scanpaths produced using different baselines and our PAT method. Our predicted scanpaths more closely resemble those of pathologists and exhibit stronger spatial correlation with tumor regions from the segmentation annotations compared to the baseline methods.
}
\label{fig:qual_scanpath}
\vspace{-2.5mm}
\end{figure*}

In Table~\ref{tab:nextfix_tab}, we compare the 5-fold cross-validation performance of the different baseline models with our models on 25 test H\&E WSIs at different magnification levels for the task of predicting the next viewport fixation in the scanpath. While the vanilla transformer models yielded the smallest Euclidean distance for the predicted next viewport fixation location and higher token similarity values for the predicted next viewport fixation, they suffer from the inability to predict magnification changes and this limitation renders these model unsuitable for predicting scanpaths (see Table \ref{tab:scanpath_tab} and Figure \ref{fig:qual_scanpath}). Our \textit{PAT} methods perform significantly better than the chance baselines in terms of the Mean-Squared-Error of the predicted viewport fixation location, although the magnification change accuracy remains low.

In Table~\ref{tab:scanpath_tab}, we compare the 5-fold cross-validation performance of the different baseline models with our models on 25 test H\&E WSIs for the scanpath prediction task given only the WSI as an input. Not only did our ``PAT-ProbMag" model outperform all baseline models, it also outperformed our prior sample based ``PAT-PriorMag" model based on the overall token similarity (TokSimScan) and NSS scores while having comparable  performance in terms of AUC. Also, we see that the prediction performance of the PAT-ProbMag model trained using Kang-PAT-H features is significantly improved compared to that produced using the DINO-PAT-H features.

\begin{table*}[t]
\centering
\setlength{\tabcolsep}{4pt}
\resizebox{\textwidth}{!}{%
\begin{tabular}{l|c|c|cccccc|cccccc}
\toprule
\textbf{Method} & \textbf{Spatial MSE $\downarrow$} & \textbf{TokSimFix $\uparrow$} & \multicolumn{6}{c|}{\textbf{Magnification Accuracy (\%) $\uparrow$}} & \multicolumn{6}{c}{\textbf{Magnification Change Accuracy (\%) $\uparrow$}} \\
\cmidrule(lr){4-9} \cmidrule(lr){10-15}
& & & \textbf{Overall} & \textbf{1X} & \textbf{2X} & \textbf{4X} & \textbf{10X} & \textbf{20X} & \textbf{Overall} & \textbf{1X} & \textbf{2X} & \textbf{4X} & \textbf{10X} & \textbf{20X} \\
\midrule
Random1 & $0.46 \pm 0.01$ & $0.39 \pm 0.02$ & $19.5 \pm 0.58$ & $18.6 \pm 1.80$ & $20.8 \pm 2.20$ & $18.9 \pm 0.48$ & $19.1 \pm 0.79$ & $19.9 \pm 1.19$ & $19.4 \pm 0.65$ & $16.7 \pm 6.12$ & $20.4 \pm 2.94$ & $19.5 \pm 1.40$ & $19.4 \pm 1.62$ & $20.0 \pm 3.28$ \\
Random2 & $0.36 \pm 0.00$ & $0.49 \pm 0.03$ & $32.9 \pm 1.71$ & $26.6 \pm 5.31$ & $24.1 \pm 3.52$ & $35.7 \pm 3.38$ & $40.1 \pm 1.27$ & $7.8 \pm 3.02$ & $\mathbf{29.9} \pm \mathbf{3.16}$ & $15.9 \pm 5.60$ & $\mathbf{24.9} \pm \mathbf{6.92}$ & $\mathbf{35.7} \pm \mathbf{4.04}$ & $\mathbf{38.5} \pm \mathbf{2.47}$ & $7.4 \pm 1.74$ \\
VanFormer & $0.07 \pm 0.00$ & $\mathbf{0.81} \pm \mathbf{0.01}$ & $\mathbf{71.4} \pm \mathbf{1.53}$ & $\mathbf{85.9} \pm \mathbf{0.78}$ & $72.2 \pm 1.85$ & $\mathbf{73.0} \pm \mathbf{0.94}$ & $72.9 \pm 2.11$ & $43.2 \pm 4.83$ & $0.0 \pm 0.0$ & $0.0 \pm 0.0$ & $0.0 \pm 0.0$ & $0.0 \pm 0.0$ & $0.0 \pm 0.0$ & $0.0 \pm 0.0$ \\
VanSemFormer (DINO-PAT-H) & $\mathbf{0.07} \pm \mathbf{0.01}$ & $0.80 \pm 0.02$ & $70.0 \pm 2.10$ & $48.1 \pm 4.91$ & $\mathbf{76.5} \pm \mathbf{3.27}$ & $72.9 \pm 2.15$ & $\mathbf{74.3} \pm \mathbf{1.32}$ & $47.1 \pm 3.92$ & $4.4 \pm 0.99$ & $3.0 \pm 2.01$ & $18.2 \pm 7.20$ & $0.3 \pm 0.12$ & $3.0 \pm 2.11$ & $5.8 \pm 1.93$ \\
PAT-ProbMag (DINO-PAT-H) & $0.16 \pm 0.01$ & $0.71 \pm 0.03$ & $57.8 \pm 1.12$ & $72.8 \pm 7.56$ & $57.8 \pm 1.12$ & $58.0 \pm 1.85$ & $58.8 \pm 1.85$ & $41.7 \pm 4.49$ & $14.9 \pm 1.56$ & $15.7 \pm 9.39$ & $11.7 \pm 3.91$  & $11.4 \pm 3.34$ & $12.7 \pm 4.92$ & $\mathbf{21.4} \pm \mathbf{3.37}$ \\
PAT-ProbMag (Kang-Vanilla) & $0.15 \pm 0.02$ & $0.68 \pm 0.04$ & $58.3 \pm 1.93$ & $72.7 \pm 7.83$ & $58.9 \pm 1.48$ & $60.6 \pm 3.11$ & $55.6 \pm 6.51$ & $43.6 \pm 1.93$ & $14.1 \pm 2.02$ & $11.4 \pm 4.59$ & $11.8 \pm 2.28$ & $11.4 \pm 3.92$ & $11.7 \pm 3.69$ & $21.4 \pm 3.96$ \\
PAT-PriorMag (Kang-PAT-H) & $0.15 \pm 0.03$ & $0.68 \pm 0.01$ & $69.5 \pm 2.30$ & $85.3 \pm 1.95$ & $69.8 \pm 3.20$ & $72.1 \pm 2.70$ & $72.1 \pm 5.40$ & $\mathbf{47.8} \pm \mathbf{3.37}$ & $4.5 \pm 1.77$ & $1.8 \pm 1.50$ & $4.5 \pm 6.29$ & $2.4 \pm 2.34$ & $0.76 \pm 1.04$ & $13.4 \pm 7.25$ \\
PAT-ProbMag (Kang-PAT-H) & $0.16 \pm 0.03$ & $0.61 \pm 0.01$ & $59.3 \pm 1.04$ & $74.8 \pm 4.39$ & $59.0 \pm 1.47$ & $59.4 \pm 1.91$ & $60.0 \pm 4.52$ & $43.1 \pm 1.26$ & $14.5 \pm 1.26$ & $\mathbf{20.4} \pm \mathbf{10.7}$ & $9.2 \pm 3.11$ & $12.8 \pm 2.15$ & $9.6 \pm 3.72$ & $20.6 \pm 4.56$ \\
\bottomrule
\end{tabular}%
}
\caption{Prediction performance on the next viewport prediction task using 5-fold cross-validation. While our PAT models do not produce the best performance for this intermediate task, they outperform chance models. The VanFormer model, closely followed by VanSemFormer, produces the best performance in terms of location MSE, TokSimFix, and overall magnification prediction accuracy, although
this good performance was due largely to the model learning to predict that the next most
probable magnification level is the same as the one from the immediately previous fixation. However, both models fail significantly at predicting magnification changes by a large margin compared to our ``PAT-PriorMag" and ``PAT-ProbMag" models, and this failure leads to its poor performance on the task of predicting scanpaths, as evidenced in Table \ref{tab:scanpath_tab} and Figure~\ref{fig:qual_scanpath}.}
\label{tab:nextfix_tab}
\end{table*}

\begin{table*}[t]
\centering
\setlength{\tabcolsep}{3pt}
\resizebox{\textwidth}{!}{%
\begin{tabular}{l|cccccc|c|c}
\toprule
\textbf{Method} & \multicolumn{6}{c|}{\textbf{Token Similarity (TokSimScan) $\uparrow$}} & \textbf{NSS $\uparrow$} & \textbf{AUC $\uparrow$} \\
\cmidrule{2-7}
 & \textbf{Overall} & \textbf{1X} & \textbf{2X} & \textbf{4X} & \textbf{10X} & \textbf{20X} & & \\
\midrule
Random1  & $0.62 \pm 0.03$ & $0.78 \pm 0.01$ & $0.77 \pm 0.01$ & $0.57 \pm 0.13$ & $0.56 \pm 0.05$ & $0.58 \pm 0.00$ & $0.05 \pm 0.00$ & $0.52 \pm 0.00$ \\
Random2  & $0.63 \pm 0.02$ & $0.91 \pm 0.01$ & \textbf{0.87 $\pm$ 0.01} & $0.67 \pm 0.12$ & $0.64 \pm 0.03$ & $0.69 \pm 0.04$ & $0.42 \pm 0.17$ & $0.67 \pm 0.04$ \\
VanFormer & $0.07 \pm 0.01$ & \textbf{0.92 $\pm$ 0.01} & $0.00 \pm 0.00$ & $0.00 \pm 0.00$ & $0.00 \pm 0.00$ & $0.00 \pm 0.00$ & $0.41 \pm 0.08$ & $0.62 \pm 0.04$ \\
VanSemFormer (DINO-PAT-H) & $0.12 \pm 0.03$ & $0.87 \pm 0.01$ & $0.79 \pm 0.03$ & $0.00 \pm 0.00$ & $0.00 \pm 0.00$ & $0.00 \pm 0.00$ & $0.16 \pm 0.08$ & $0.60 \pm 0.01$ \\
GazeFormer (DINO-PAT-H)  & $0.67 \pm 0.00$ & $0.74 \pm 0.02$ & $0.81 \pm 0.00$ & $0.70 \pm 0.00$ & $0.55 \pm 0.02$ & $0.66 \pm 0.01$ & $0.17 \pm 0.05$ & $0.63 \pm 0.01$ \\
PAT-ProbMag (DINO-PAT-H) & $0.73 \pm 0.04$ & $0.71 \pm 0.06$ & $0.76 \pm 0.08$ & $0.76 \pm 0.04$ & $0.72 \pm 0.04$ & $ 0.72\pm 0.04$ & $0.83 \pm 0.13$ & $0.71 \pm 0.04$ \\
PAT-ProbMag (Kang-Vanilla) & $0.71 \pm 0.04$ & $0.70 \pm 0.06$ & $0.73 \pm 0.03$ & $0.73 \pm 0.04$ & $0.69 \pm 0.04$ & $0.68 \pm 0.04$ & $0.97 \pm 0.10$ & \textbf{0.75 $\pm$ 0.11} \\
PAT-PriorMag (Kang-PAT-H) & \textbf{0.80 $\pm$ 0.01} & $0.81 \pm 0.07$ & $0.80 \pm 0.06$ & \textbf{0.84 $\pm$ 0.06} & $0.72 \pm 0.05$ & \textbf{0.82 $\pm$ 0.05} & $0.97 \pm 0.11$ & 0.74 $\pm$ 0.02 \\
PAT-ProbMag (Kang-PAT-H) & \textbf{0.80 $\pm$ 0.03} & 0.78 $\pm$ 0.01 & 0.80 $\pm$ 0.12 & 0.83 $\pm$ 0.06 & \textbf{0.77 $\pm$ 0.10} & 0.81 $\pm$ 0.08 & \textbf{0.99 $\pm$ 0.10} & 0.74 $\pm$ 0.02 \\
\bottomrule
\end{tabular}%
}
\caption{Quantitative evaluation of the prediction performance of our PAT models using 5-fold cross-validation. Our \textit{PAT-ProbMag} model outperforms other baselines in the model comparison.}
\label{tab:scanpath_tab}
\end{table*}

\begin{table}[t]
    \setlength{\tabcolsep}{2pt}
        \centering
        \resizebox{0.65\textwidth}{!}{
            \begin{tabular}{l|c}
                \toprule
                \textbf{Method} & Semantic Sequence Score (SSS)
                 \\
                \midrule
                Human & 0.420 \\
                Random1   & 0.427  \\
                Random2 & 0.364 \\
                VanFormer & 0.412 \\
                VanSemFormer (DINO-PAT-H) & 0.376 \\
                GazeFormer (DINO-PAT-H) &  0.366 \\
                PAT-PriorMag (DINO-PAT-H) & 0.461 \\
                PAT-ProbMag (DINO-PAT-H) &  \textbf{0.467} \\
                \bottomrule
            \end{tabular}
        }

    \caption{Comparison of the Semantic Sequence Score (SSS) metric for our proposed PAT model with different baseline models on 13 test WSIs. Gleason grade segmentations used to compute SSS were provided by a GU specialist. Our \textit{PAT-ProbMag} model outperforms the other baselines.}
    \label{tab:sss_comp}
\end{table}

\begin{figure*}
\centering
\includegraphics[width = 13.80cm]{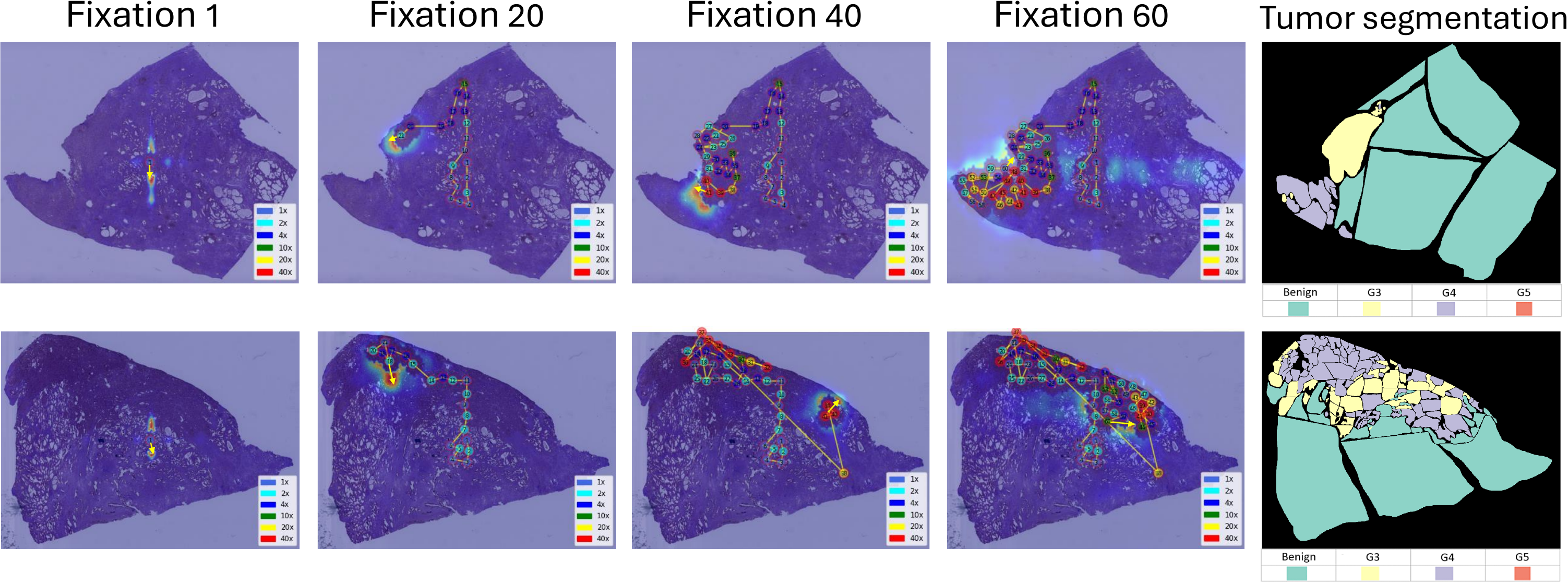}
\caption {Depiction of how predictions from our PAT model evolve over time for two WSIs. Note the convergence of attention over time to the highest tumor grades. 
}
\label{fig:qual_scanevol}
\end{figure*}

In Table~\ref{tab:sss_comp}, we compare the Semantic Sequence Score (SSS) of our proposed PAT model against several baselines on 13 test WSIs annotated with Gleason-grade segmentations by a genitourinary (GU) pathology specialist. The \textit{PAT-ProbMag} variant achieves the highest SSS, indicating its superior ability to predict scanpaths that align with clinically meaningful transitions across regions with different Gleason patterns. To contextualize this performance, we also measured the inter-pathologist agreement by computing pairwise SSS between individual pathologist scanpaths, which yielded a relatively low average of 0.420. This reflects substantial inter-observer variability in attention during WSI reading—a known challenge in pathology. Despite this inherent variability in the training data, our model learns consistent attention patterns across pathologists, leading to better  semantic alignment of the predicted scanpaths with the pathologist-derived scanpaths and an improved attention prediction performance. 

Figure~\ref{fig:qual_scanpath} shows a qualitative comparison of the scanpaths predicted by several baseline models with those from our proposed models. Ground-truth scanpaths from pathologists are also shown. The \textit{Random1} baseline randomly allocates fixations across the WSI, as expected. Although the \textit{Random2} baseline is derived from a pathologist’s scanpath, it originates from a different WSI and therefore fails to accurately explore tumor regions. The \textit{VanFormer} and \textit{VanSemFormer} baselines are inaccurate in their prediction of very small inter-fixation distances and fail to make significant changes in magnification. The \textit{VanFormer} model produces identical scanpaths, always scanning out from the center of a WSI at 1X magnification, regardless of the input image. In contrast, the \textit{VanSemFormer} model generates scanpaths that vary based on the input WSI by considering token information across different magnifications. However, it also suffers from overly short fixation shifts and too few magnification changes, typically transitioning only from 1X to 2X. The \textit{GazeFormer} model, due to its non-autoregressive nature, predicts the entire scanpath in a single step. This approach, as observed, often fails to produce accurate scanpaths. Our \textit{PAT} models, both prior sample based (\textit{PAT-PriorMag}) and probabilistic (\textit{PAT-ProbMag}), more closely resemble the pathologist scanpaths. They also better cover the tumor regions compared to the baselines, based on tumor segmentation annotations obtained from a GU specialist. However, the magnification transitions in the \textit{PAT-PriorMag} model are overly uniform and fail to capture the variability in a pathologist’s attention, resulting in unrealistic transitions. In contrast, our \textit{PAT-ProbMag} model addresses this limitation by learning magnification transitions in a probabilistic manner during training, leading to more realistic scanpaths.

\begin{figure*}
\centering
\includegraphics[width = 13.80cm]{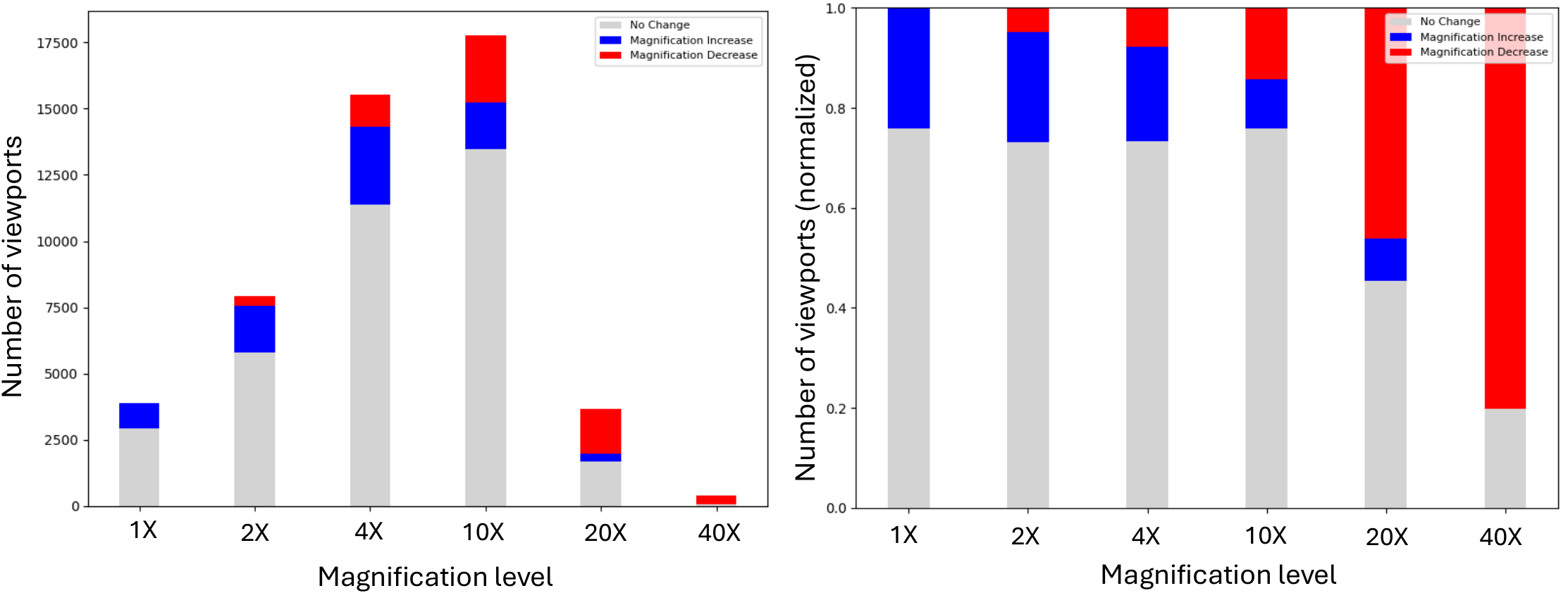}
\caption {Magnification transition statistics across six different magnification levels. Transitions from lower magnifications (1X, 2X, and 4X) to higher magnifications (zooming in) are common, whereas transitions from higher magnifications (10X, 20X, and 40X) to lower magnifications (zooming out) frequently occur.
}
\label{fig:mag_transit}
\end{figure*}

In Figure~\ref{fig:qual_scanevol}, we depict the temporal evolution of our predicted attention scanpaths for two test WSIs \textit{TCGA-EJ-7315} and \textit{TCGA-EJ-7784} from the TCGA-PRAD dataset. Specifically, for a viewport fixation at time step $N$, we visualize the predicted attention heatmap, $\hat{H}_{N+1}$ that decides the location of the next viewport fixation $V_{N+1}$, conditioned on the history of previous viewport fixations $\{V\}_{n=1}^N$, as well as the corresponding attention scanpath, $S_N$. The arrow after each predicted last viewport fixation indicates the location of maximum intensity in the predicted heatmap, where the next viewport fixation would be selected. We observe that the predicted scanpath trajectories tend to converge toward the tumor regions (based on the tumor segmentation map), originating from the center of the WSI.

We also compared the attention heatmap prediction performance of our PAT-H sub-network with several baseline models, including Frozen ResNet50 and DINO backbones, as well as prior attention modeling approaches such as ProstAttNet \citep{chakraborty2022visual} and PathAttFormer \citep{chakraborty2022predicting}. Across all magnification levels (2X, 4X, 10X, and 20X), our PAT-H model—especially when using Kang et al. features—consistently outperformed baselines. Notably, PAT-H (w/Kang) achieved the best scores at each magnification, demonstrating a strong ability to align with ground-truth attention distributions. For example, at 10X, PAT-H (w/Kang) reached a CC of 0.765 and an NSS of 2.223, substantially higher than prior state-of-the-art methods, highlighting the benefit of multi-resolution modeling and domain-specific feature selection in attention prediction for digital pathology. See the supplementary material for more details.

\subsection{Ablation studies}
\label{sec:ab_study}
\textbf{Magnification transition frequency.} Figure~\ref{fig:mag_transit} visualizes the frequency of transitions in magnification levels, both in terms of number of viewport transitions at a given magnification (left) and the normalized version of the same (right). The stacked bars indicate how often pathologists maintain their magnification level (no change), increase, or decrease it while navigating between different magnifications.

From the left plot, we observe that 10X is the most frequently used magnification level, followed by 4X and 2X. As expected, when a pathologist is at a relatively low magnification (e.g. 1X, 2X) there is a higher proportion of changes to a higher magnification. However, we also observed significant periods of low-magnification scanning, likely indicating an initial exploration phase where pathologists decide where to zoom in to examine regions in more detail. In contrast, higher magnifications (20X, 40X) primarily show no changes or decreases in magnification, also as expected. For example, magnification changes at 4X mostly lead to an increase in magnification (to 10X or higher), whereas changes while at 10X more frequently lead to a decrease in magnification (to 4X or lower). These data patterns support the use of lower magnifications for initial exploration followed by the use of higher magnifications for more detailed.

\textbf{Feature Encodings.} We evaluated the predictive performance of our model using different types of feature encodings: DINO-Vanilla, DINO-PAT-H, Kang-Vanilla, and Kang-PAT-H. We found that feature encodings derived from our \textit{PAT-H} sub-network consistently improved performance across all metrics. Detailed ablation results are provided in the supplementary material. This highlights the superior ability of our model's learned features to capture pathologist attention patterns compared to vanilla DINO and Kang features.

\textbf{Feature resolution.} We also ablated our model across multi-resolutional feature encodings derived from different magnification levels. We observed that 10X magnification produced the best high-resolution feature space for predicting scanpaths, likely because it is the most commonly used magnification level during WSI reading. Detailed results are provided in the supplementary material.

%% file: sec/conclusions_6.tex
\section{Conclusion}
\label{sec:conclusion_6}

We present a two-stage model to predict the dynamic attention of pathologists as they read WSIs of prostate cancer for grading. By tracking their viewport movements during WSI reading, we gathered attention data from 43 pathologists over 123 WSIs. Employing transformer-based models, we predicted the attention scanpaths of pathologists, achieving levels of performance surpassing chance and baseline models.  

Our method can be used to provide feedback to trainee pathologists on where and when in a WSI to allocate their visual attention, thus teaching them how to view and grade WSIs like an expert. Our model can also be integrated into decision support and training systems to guide pathologists during image assessment. For instance, as a trainee navigates a WSI, the system might highlight regions that an expert would likely examine, suggesting optimal magnifications and traversal sequences. This guidance has the potential to help in identifying critical diagnostic features that might otherwise be overlooked, thereby enhancing diagnostic accuracy and efficiency. We believe that this will be crucial for pathology training and competency assessment, offering a pathway to enhance grading consensus among non-specialists by emulating AI specialists' attention patterns.

While we acknowledge that testing across multiple cancer types is necessary for full validation, the combination of a large, diverse dataset and a scalable modeling approach makes this work an important step towards broader applicability in digital pathology. Future work will involve testing the model’s effectiveness across multiple cancer types and pathology subspecialties to further establish its generalizability. Additionally, in ongoing work, we are attempting to further improve our attention predictions by using explicit semantic information as a model input. Such information could be encoded in the form of semantic segmentation maps that capture the presence of factors that are clinically significant for the task of grading WSIs of prostate cancer, such as the different Gleason patterns (such as Benign/G3/G4/G5) \citep{bulten2020automated}, cribriform pattern \citep{ambrosini2020automated} (a strong indicator of the presence of G4 grade tumor), and various other patterns and glandular abnormalities that are standardized on clinical pathology reports. We hypothesize that using such specialized information explicitly will significantly improve the performance of our predictive models.

%% file: sec/acknowledgments_7.tex
\section{Acknowledgments}
\label{sec:acknow}
\noindent This work was supported by a seed grant from the Stony Brook University Office of the Vice President for Research (1150956-3-63845), NSF grants IIS-2212046 and IIS-2123920, and grants UH3-CA225021, U24-CA215109, and U24-CA180924 from the NCI and NIH.